\documentclass[pra,twocolumn,showpacs,floatfix,superscriptaddress,aps]{revtex4}
\usepackage{amsmath}
\usepackage{makeidx}
\usepackage{amssymb}
\usepackage{graphicx}

\usepackage[usenames]{color}
\usepackage[normalem]{ulem}

\begin{document}

\title{Stability and collapse of fermions in a binary dipolar boson-fermion $^{164}$Dy-$^{161}$Dy mixture }


\author{S. K. Adhikari\footnote{adhikari@ift.unesp.br; URL  http://www.ift.unesp.br/users/adhikari  }} \affiliation{Instituto de F\'{\i}sica Te\'orica,
UNESP - Universidade Estadual Paulista, 01.140-070 S\~ao Paulo, S\~ao
Paulo, Brazil}

\begin{abstract} We suggest a time-dependent mean-field hydrodynamic 
model for a binary dipolar boson-fermion mixture to study the stability 
and collapse of fermions in the $^{164}$Dy-$^{161}$Dy mixture. The 
condition of stability of the dipolar mixture is illustrated in terms of 
phase diagrams. A collapse is induced in a disk-shaped stable binary 
mixture by jumping the interspecies contact interaction from repulsive to 
attractive by the Feshbach resonance technique. The subsequent dynamics 
is studied by solving the time-dependent mean-field model including 
three-body loss due to molecule formation in boson-fermion and 
boson-boson channels. Collapse and fragmentation  in the fermions
after subsequent 
explosions are illustrated. The anisotropic dipolar 
interaction leads to anisotropic fermionic density distribution during 
collapse. The present study is carried out in three-dimensional space 
using realistic values of dipolar and contact interactions.
 \end{abstract}


\maketitle
 
\section{Introduction}

The alkali metal atoms used in  most Bose-Einstein condensation
(BEC) experiments  
have negligible dipole moment. However,  BECs
of magnetically polarized atoms, for instance $^{52}$Cr \cite{pfau}, $^{168}$Er \cite{grimm}, and 
$^{164}$Dy  \cite{boselev} atoms, with reasonably large magnetic moments
 have been realized. 
Polar molecules with  much larger   electric dipole moments  are also being considered for 
BEC experiments \cite{becmol}.
Thus one can study the 
interplay between the long-range anisotropic dipolar interaction 
and a  variable  short-range interaction \cite{pfau}
using a Feshbach resonance \cite{fesh} in a dipolar BEC.  
The stability of a dipolar BEC depends not only 
on the  atomic interaction, but also strongly on  trap geometry \cite{pfau,10}.
In a disk configuration the  dipolar interaction is
repulsive and a dipolar BEC is more stable; while, in a cigar 
configuration the  dipolar interaction is
attractive and a dipolar BEC is less stable due to  
collapse instability.
A remarkable feature of the stability of a dipolar BEC is that, irrespective of the underlying trap symmetry, a dipolar BEC always becomes unstable to 
collapse
with 
the increase of  dipolar interaction or of the number of atoms \cite{cola,colb,dipolarb}. This means 
that, in the disk shape with predominantly repulsive   
dipolar interaction, a dipolar BEC is unstable beyond a critical number of atoms even for repulsive  short-range atomic interaction.   
This leads to peculiar stability phase diagrams for single-component \cite{cola,dipolarb}
and binary dipolar \cite{dipolarbin}
BECs.
The controllable
short-range interaction together with the long-range dipolar interaction 
make the dipolar BEC an attractive system for
experimental study
and a challenging system for
theoretical investigation.  Because of the 
peculiar features  there has been enhanced interest in
the study of dynamic as well as static properties of dipolar BECs.
Among the
novel features of a dipolar BEC, 
one can mention the peculiar  stability
phase diagrams  \cite{10}, red-blood-cell-like
biconcave  density distribution due to radial and angular
roton-like excitations \cite{11}, anisotropic D-wave collapse
\cite{12}, formation of anisotropic soliton, vortex soliton \cite{13}
and vortex lattice \cite{14}, anisotropic shock and sound waves \cite{15},
entanglement \cite{enlo},
localization in 
disordered potential \cite{loca}
and anisotropic Landau critical velocity
\cite{16} among others. Stable checkerboard, star, and stripe configurations in dipolar BECs have been identified 
in a two-dimensional (2D) optical lattice as stable Mott insulator \cite{17} as well as superfluid soliton \cite{18}
states. A new possibility of studying universal properties 
of dipolar BECs at unitarity has been
suggested \cite{19}.

After the pioneering experiments on BEC of alkali-metal atoms, degenerate  
spin-polarized
gases of
fermionic $^6$Li \cite{117}, $^{40}$K \cite{118}, and $^{87}$Sr \cite{119} atoms were observed. Later, superfluid states
of paired $^6$Li \cite{120} and $^{40}$K \cite{121} atoms have also been studied. More recently a
degenerate dipolar gas of fermionic $^{161}$Dy atoms with large magnetic moment
has been created and studied \cite{122}.  Fermionic polar molecules, such as $^{40}$K$^{87}$Rb,  are also being
considered for these studies \cite{123}.
With a dipole moment of 0.6 Debye,
 the $^{40}$K$^{87}$Rb molecule in the singlet
rovibrational ground state has  a dipolar interaction larger than in
 $^{161}$Dy atoms by more than an order
of magnitude \cite{123,rep}.  A trapped degenerate spin-polarized 
nondipolar fermionic gas 
is absolutely stable for any number of atoms owing to strong Pauli repulsion among identical fermions. However, due to the peculiar 
nature of dipolar interaction, a trapped 
degenerate dipolar spin-polarized 
fermionic gas is unstable for the number of atoms above a 
critical value \cite{dipolarb,pucol}. Although, the short-range $S$-wave contact interaction is absent in nondipolar fermionic gas, the dipolar interaction operative in the fermionic gas is responsible for the collapse instability.

Collapse dynamics in a nondipolar BEC of $^{85}$Rb atoms, initiated by 
jumping the scattering length from positive (repulsive) to negative 
(attractive) near a Feshbach resonance \cite{fesh}, has revealed many 
fascinating novel features \cite{donley,13611,thcoll}, such as, jet formation 
and explosion after collapse. Although collapse is forbidden in a gas of 
degenerate trapped single-component nondipolar fermions in 
isolation due to Pauli repulsion among identical fermions, it is 
possible in a trapped degenerate boson-fermion mixture with attractive 
short-range boson-fermion interaction. There have been experimental 
\cite{modugno} and theoretical \cite{thcollbf} investigations of fermion 
collapse in a binary boson-fermion $^{87}$Rb-$^{40}$K mixture. The 
presence of the bosons facilitates collapse via attractive interspecies 
interaction. 

Recently, there has been interest in creating and studying 
a highly dipolar trapped boson-fermion mixture of dysprosium isotopes 
\cite{boselev,122}. In the $^{164}$Dy-$^{161}$Dy
 boson-fermion mixture
the presence of the bosonic  $^{164}$Dy
atoms will favor the collapse in the fermionic $^{161}$Dy atoms for attractive boson-fermion contact interaction. Hence we study the statics and dynamics of collapse of fermionic $^{161}$Dy atoms in the trapped binary  $^{164}$Dy-$^{161}$Dy mixture using a mean-field model. 
The stability of the mixture is illustrated by phase diagrams showing the critical  number of   $^{164}$Dy atoms in the stable mixture for different 
values of intraspecies boson-boson and interspecies boson-fermion scattering lengths and for different trap aspect ratios. To study the dynamics of collapse we include in our model two types of three-body loss by the formation of  boson-boson and boson-fermion molecules. In the presence of these three-body 
losses we study the evolution of the number of  $^{164}$Dy and $^{161}$Dy
 atoms during collapse.  After the initial quick loss of a large number of atoms  
quasi-stable bosonic and fermionic remnants  are formed, which  
are found to last for a large interval of time. The three-dimensional (3D) isodensity contours of the remnants clearly exhibit the dynamics of fragmentation. The non-$S$ wave nature of the dipolar interaction is clearly seen in the 2D and 3D isodensity contours.

In Sec. II we present the mean-field model for studying the statics and dynamics of the binary dipolar boson-fermion $^{164}$Dy-$^{161}$Dy
mixture.
In this model we incorporate all interspecies and intraspecies short-range $S$-wave and long-range dipolar interactions.
Two possibilities of three-body loss by the formation of boson-boson $^{164}$Dy$^{164}$Dy and boson-fermion $^{164}$Dy$^{161}$Dy 
molecules are also included in the dynamics. In Sec. III we report the results of numerical investigation. The stability of the system 
is demonstrated in terms phase diagrams. We also present the results of dynamical evolution of the system during collapse and illustrate 
non-$S$-wave density distribution during collapse due to dipolar interaction. Finally, in Sec. IV we present a brief summary of
the present study and future perspectives.

\section{Mean-field hydrodynamic model}

 {
We  study the degenerate mixture of $N_b$ bosonic atoms of mass $m_b$ and 
$N_f$ spin-up fermionic atoms of mass $m_f$ under the action of interspecies and intraspecies  isotropic
short-range  and anisotropic dipolar  interactions at zero temperature. The bosonic atoms 
form a BEC and superfluid $p$-wave pairing in the fermionic system occurs for dipolar interactions 
beyond a critical strength for a definite trap aspect ratio \cite{fs}.
 In the presence of superfluid pairing the bosonic and fermionic superfluids are
described by the order parameters $\Phi_b$ and $\Phi_f$, respectively.
In this hydrodynamic
description of fermionic superfluid we neglect
the gap and take the density $n_f = |\Phi_f|^2,$ although the 
gap can be accommodated in an improved hydrodynamic
model \cite{st}.
Assuming that the superfluidity in fermions has been achieved we present 
the following mean-field model for the binary superfluid. However, if the 
parameters of the model, e.g., number of atoms, dipolar strength, trap aspect ratio etc., are below the critical limit for attaining pairing in fermions,
the present model will still be valid where  $\Phi_f$ will not be 
the fermionic order parameter but should be related to the density $n_f$   via $n_f =|\Phi_f|^2$. { The fermionic system will then be in a collisional hydrodynamic regime \cite{axel}
where the hydrodynamic behavior appears 
due to collision among the dipolar fermions. Unless we study rotational properties, e.g., quantized vortex formation, collisional hydrodynamic and superfluid phases cannot be distinguished easily and will be described by the same hydrodynamics equations.}
A complete analysis to determine the critical limit of attaining fermion pairing in the binary dipolar boson-fermion mixture is beyond the scope of this study and  here we only relate 
the function $\Phi_f$ to density and not to any other superfluid property
of fermions.  A similar interpretation of the fermionic function  $\Phi_f$ in 
the normal fermionic gas has led a satisfactory description of the nondipolar 
boson-fermion mixture \cite{ndbf,thcollbf}.

The fermionic atoms  are treated by a hydrodynamical 
Lagrangian and the bosonic atoms by the mean-field 
Gross-Pitaevskii Lagrangian.
After including the dipolar interaction terms following Refs. \cite{dipolardrop,dipolarbin} in
 the standard Lagrangian density of the boson-fermion mixture \cite{adhibf,extra},
the Euler-Lagrange equations for the binary mixture 
can be written as  
\begin{align}  \label{gp3db} 
&i\hbar   \frac{\partial \Phi_b({\bf r},t)}{\partial t}
 =  \biggr[ -\frac{\hbar^2}{2m_b}\nabla^2  + \frac{1}{2}m_b\omega_b^2 \biggr(\frac{\rho^2}{\lambda_b^{2/3}}
+z^2\lambda_b^{4/3}\biggr)\nonumber \\
&+\frac{4\hbar^2\pi a_b }{m_b}|\Phi_b|^2
+\frac{2\pi \hbar^2}{m_R}a_{bf}|\Phi_f|^2\nonumber \\
&+ \frac{ \mu_0 \mu_b^2}{4\pi}\int V_{dd}({\bf r -r'})|\Phi_b({\bf r'})|^2d{\bf r'}\nonumber \\
&+ \frac{ \mu_0 \mu_b\mu_f}{4\pi}\int V_{dd}({\bf r -r'})|\Phi_f({\bf r'})|^2d{\bf r'}
\biggr] \Phi_b({\bf r},t), \\
&i\hbar   \frac{\partial \Phi_f({\bf r},t)}{\partial t}
 =  \biggr[ -\frac{\hbar^2}{8m_f}\nabla^2   + \frac{1}{2}m_f\omega_f^2 \biggr(\frac{\rho^2}{\lambda_f^{2/3}}
+z^2\lambda_f^{4/3}\biggr)
\nonumber \\
& +\frac{ \mu_0 \mu_b\mu_f}{4\pi}\int V_{dd}({\bf r -r'})|\Phi_b({\bf r'})|^2d{\bf r'}
+\frac{2\pi \hbar^2}{m_R}a_{bf}|\Phi_b|^2
\nonumber \\
&
+ \frac{ \mu_0 \mu_f^2}{4\pi}\int V_{dd}({\bf r -r'})|\Phi_f({\bf r'})|^2d{\bf r'}+\mu_F
\biggr] \Phi_f({\bf r},t),\label{gp3df}
\end{align}
with normalization $\int |\Phi_i({\bf r})|^2 d{\bf r} = N_i, i=b,f,$ 
where we have included the dipolar interaction terms following Ref. \cite{dipolarbin}.
{ The  prefactor  $-\hbar^2/8m_f$ in the space derivative term of the   
 fermionic equation 
(\ref{gp3df})  takes into account the possibility of pairing in dipolar fermions  and leads to a Galilei-invariant dynamics \cite{adhibf}.  The space derivative terms have a quantum origin and are called the quantum pressure terms.
In the absence of these terms Eqs. (\ref{gp3db}) and (\ref{gp3df}) reduce to the classical hydrodynamic equations, while the fermionic equation (\ref{gp3df}) becomes the well-known local density approximation. In the absence of pairing in the normal state the prefactor $-\hbar^2/8m_f$ was suggested by von Weizs\"acker for a proper description of density \cite{weiz}.}
Here $\mu_0$ is the permeability of free space, $\mu_b$ and $\mu_f$ are magnetic moments of bosons and fermions, respectively,
$
 V_{dd}({\bf R}) = (1-3\cos^2\theta)$ $/R^3,$ $\quad 
{\bf R=r-r'},
$ $ 
m_{R}= m_b m_f/(m_b+m_f)$, and $\theta$ is the angle between the vector 
$\bf R$ and the polarization direction $z$. The traps are axially symmetric
for bosons and fermions with average angular 
frequencies $\omega_i$ and aspect ratios $\lambda_i=\omega_{zi}/\omega_{\rho i}$  with $z$ and $\rho$ denoting axial  (polarization) and transverse directions, respectively: $\omega_i=(\omega_{\rho i}^2\omega_{z i})^{1/3}$.  
In Eq. (\ref{gp3df}) $\mu_F$ is the bulk chemical potential of the free 
spin-polarized
fermionic gas  
\begin{equation}
\mu_F =\frac{\hbar^2(6\pi^2   |\Phi_f|^2)^{2/3}}{2m_f}.
\end{equation}
Equations (\ref{gp3db}) and (\ref{gp3df}) are essentially the same as similar equations for a binary bosonic dipolar mixture \cite{dipolarbin} with the intraspecies contact interaction in the second bosonic component replaced by the fermionic bulk chemical potential $\mu_F$.


To study the collapse dynamics in the dipolar boson-fermion mixture 
we have to add
the mechanisms for atom loss in Eqs. (\ref{gp3db}) and
(\ref{gp3df}). 
The net interspecies boson-fermion and boson-boson
interactions  lead to the following three-body
recombination processes to form a boson-fermion (BF) and boson-boson (BB)
molecules responsible
for the loss of atoms \cite{modugno}
\begin{eqnarray}
B + B + F \to  (BF) + B, \label{bfmol} \\
B + B + F \to  (BB) + F. \label{xx}
\end{eqnarray}
In addition, there is also the
possibility of the formation of a boson-boson molecule
by the reaction
\begin{equation}\label{bbmol}
B + B + B \to (BB) + B. 
\end{equation}
Although, not prohibitive in the presence of dipolar interaction, 
we neglect the formation of two-fermion molecule due to strong Pauli 
repulsion among identical spin-polarized fermions. For the same reason we also 
neglect the three-body loss initiated by a boson and two fermions or by three fermions.
Consequently, fermionic atoms could only be lost in the presence
of bosons according to reaction (\ref{bfmol}) and the loss rate scales quadratically with
bosonic density and is independent of fermion number
$N_f$ \cite{modugno}. There is also loss of bosonic atoms due to reactions 
(\ref{bfmol})  $-$
(\ref{bbmol}). These reactions will contribute to imaginary (dissipative) loss terms in  Eqs. (\ref{gp3db}) and
(\ref{gp3df}). 
Of these, reaction (\ref{bbmol}) contributes to the loss term 
$-i\hbar K_3^{(bb)} |\Phi_b|^4/2$ in Eq. (\ref{gp3db}), reaction  (\ref{bfmol}) contributes to the loss term 
$-i\hbar K_3^{(bf)} |\Phi_b|^2|\Phi_f|^2/2$ in Eq. (\ref{gp3db})  and
$-i\hbar K_3^{(bf)} |\Phi_b|^4/2$ in Eq. (\ref{gp3df}), and  reaction 
(\ref{xx}) contributes to a loss term 
$-i\hbar K_3^{(bbf)} |\Phi_b|^2|\Phi_f|^2/2$ in Eq. (\ref{gp3db}), where  $K_3^{(bf)}$,  $K_3^{(bbf)}$, and $K_3^{(bb)}$ are the respective loss rates of the 
reactions  (\ref{bfmol}),  (\ref{xx}),  and (\ref{bbmol}), respectively.
As the main interest of this study is to investigate the fermionic collapse, 
and as the experimental loss rates are not yet known, we combine contributions of bosonic loss terms of reactions  (\ref{bfmol}) and (\ref{xx}) into a single term  in Eq. (\ref{gp3db}).


To compare the dipolar and contact interactions, the intra- and interspecies dipolar interactions will be  expressed in terms the lengths
$a^{(i)}_{dd}$ ($i=b,f$) and $a^{(bf)}_{dd}$, respectively, defined by
\begin{align}
\frac{\mu_0\mu_i^2}{4\pi}= \frac{3\hbar^2}{m_i} a^{(i)}_{dd}, \quad
 \frac{\mu_0\mu_b\mu_f}{4\pi}= \frac{3\hbar^2}{2 m_R} a^{(bf)}_{dd}.
\end{align}
  We express the strengths of the dipolar interactions in Eqs. (\ref{gp3db}) and (\ref{gp3df}) by these dipolar lengths 
and transform these equations into the following dimensionless form:
\begin{align}& \,
i \frac{\partial \phi_b({\bf r},t)}{\partial t}=
{\Big [}  -\frac{\nabla^2}{2 }
+ \frac{1 }{2} \Big(
\frac{\rho^2}{\lambda_b^{2/3}}+\lambda^{4/3}_b z^2 \Big) 
+ g_{bf} \vert \phi_f \vert^2
\nonumber  \\ &
+ g_{dd}^{(b)}
\int V_{dd}({\mathbf R})\vert\phi_b({\mathbf r'},t)\vert^2 d{\mathbf r}'
+ g_{b} \vert \phi_b\vert^2\nonumber  \\ &
+ g_{dd}^{(bf)}
\int V_{dd}({\mathbf R})\vert\phi_f({\mathbf r'},t)\vert^2 d{\mathbf r}' 
\nonumber\\ & \, %
 -\frac{i}{2}k_3^{(bb)}N_b^2|\phi_b|^4
-\frac{i}{2}k_3^{(bf)}N_bN_f|\phi_b|^2|\phi_f|^2 
{\Big ]}  \phi_b({\bf r},t),
\label{eq3}
\end{align}
\begin{align}
& \,
i \frac{\partial \phi_f({\bf r},t)}{\partial t}={\Big [}  
-m_{bf} \frac{\nabla^2}{8}
+ \frac{m_w}{2}\Big(\frac{\rho^2}{\lambda_f^{2/3}}+\lambda^{4/3}_f z^2 \Big)\nonumber  \\ &
+ g_{dd}^{(f)}
\int V_{dd}({\mathbf R})\vert\phi_f({\mathbf r'},t)\vert^2 d{\mathbf r}'
+\frac{m_{bf}}{2}(6 \pi^2N_f|\phi_f|^2)^{2/3} \nonumber \\ & \,
+ g_{fb} \vert \phi_b \vert^2
+ g_{dd}^{(fb)}
\int V_{dd}({\mathbf R})\vert\phi_b({\mathbf r'},t)\vert^2 d{\mathbf r}' 
\nonumber  \\ &
 -\frac{i}{2}k_3^{(bf)}N_b^2|\phi_b|^4
{\Big ]}  \phi_f({\bf r},t),
\label{eq4}
\end{align}
with normalization $\int |\phi_i{(\bf r)}|^2 d{\mathbf r}=1,$
where 
$m_{bf}={m_b}/{m_f},$
$m_w={ \omega_f^2}/{(m_{bf}\omega_b^2)},$
$g_b=4\pi a_b N_b,$
$g_{dd}^{(b)}=3N_b a_{dd}^{(b)},$
$g_{bf}={2\pi m_b} a_{bf} N_f/m_R,$
$g_{fb}={2\pi m_b} a_{bf} N_b/m_R,$
$g_{dd}^{(f)}= 3N_f a_{dd}^{(f)}m_{bf},$
$g_{dd}^{(bf)}=3N_f  a_{dd}^{(bf)}(m_b/2m_R),$
$g_{dd}^{(fb)}=3N_b a_{dd}^{(bf)}(m_b/2m_R). $ 
In Eqs. (\ref{eq3}) and (\ref{eq4}), length is expressed in units of oscillator length for boson $l_0=\sqrt{\hbar/m_b\omega}$, energy 
and chemical potential 
in units of oscillator energy $\hbar\omega$, density $|\phi_i|^2$ in units of $l_0^{-3}$, and time in units of $ t_0=\omega^{-1}$, with $\omega\equiv \omega_b$,
$k_3^{(bf)}\equiv K_3^{(bf)}/(\omega_b l_0^6)$ and $k_3^{(bb)}  \equiv K_3^{(bb)}/(\omega_b l_0^6)  $ are the dimensionless three-body loss rates for the formation of 
(BF) and (BB) molecules.

\begin{figure}
\begin{center}
\includegraphics[width=\linewidth]{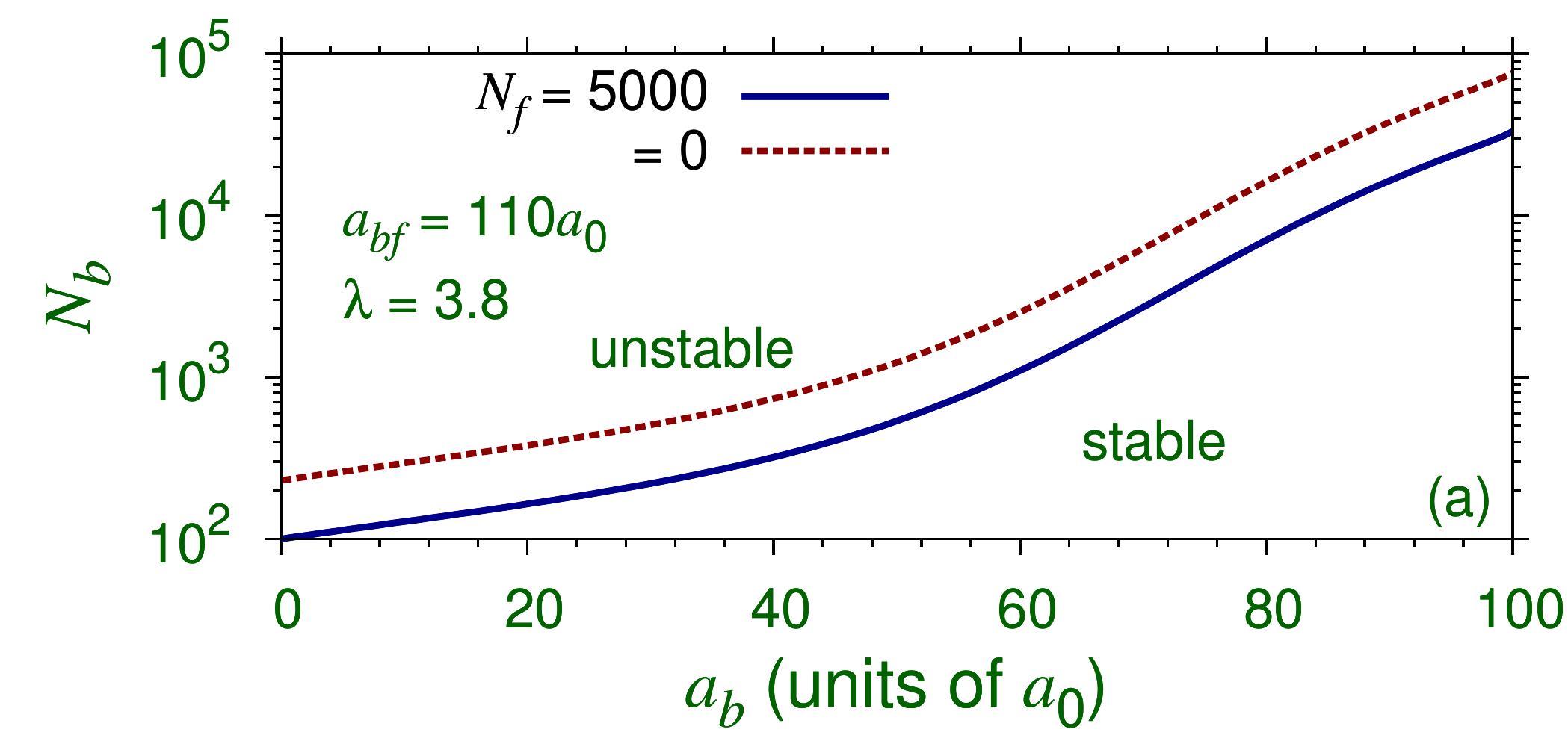}
\includegraphics[width=\linewidth]{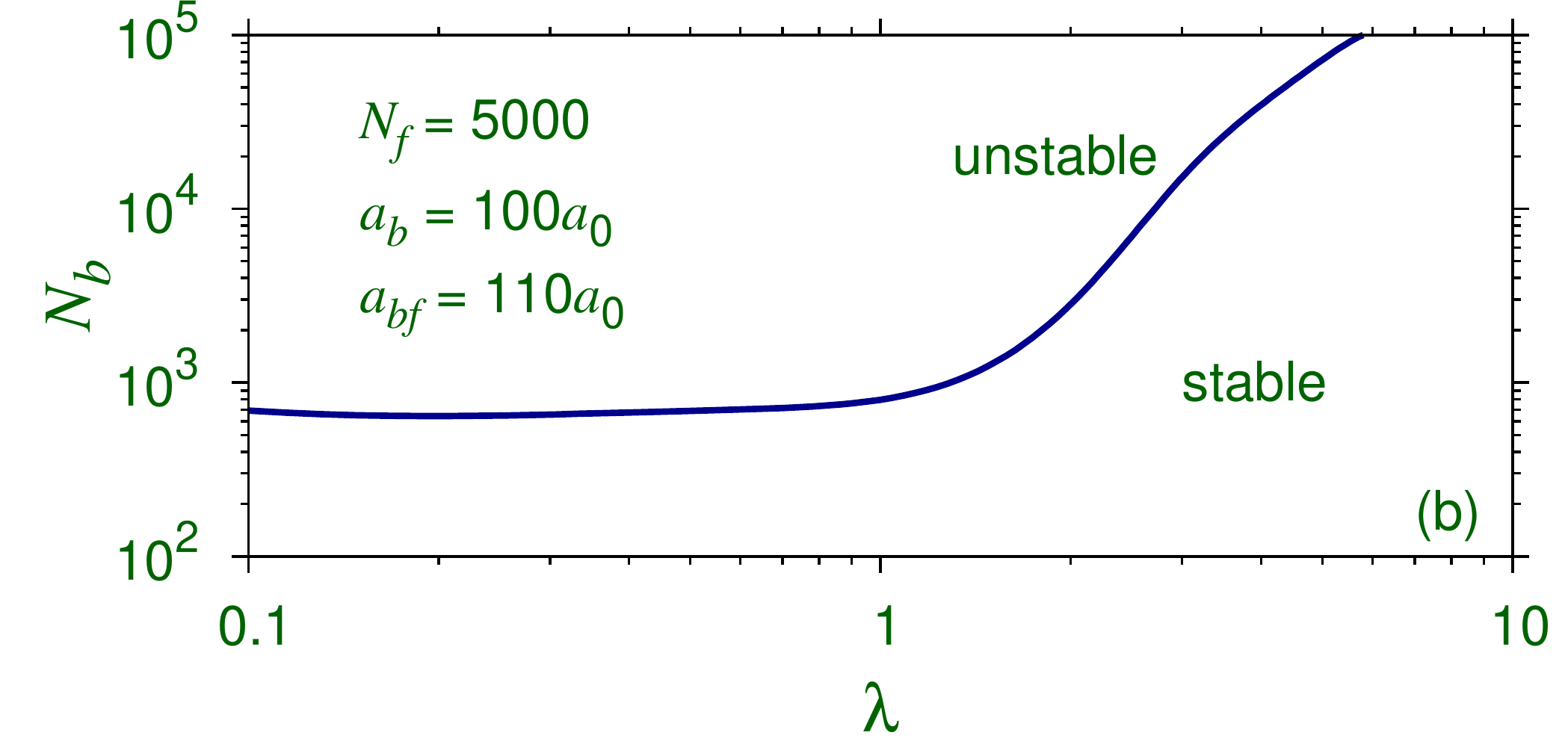}
\includegraphics[width=\linewidth]{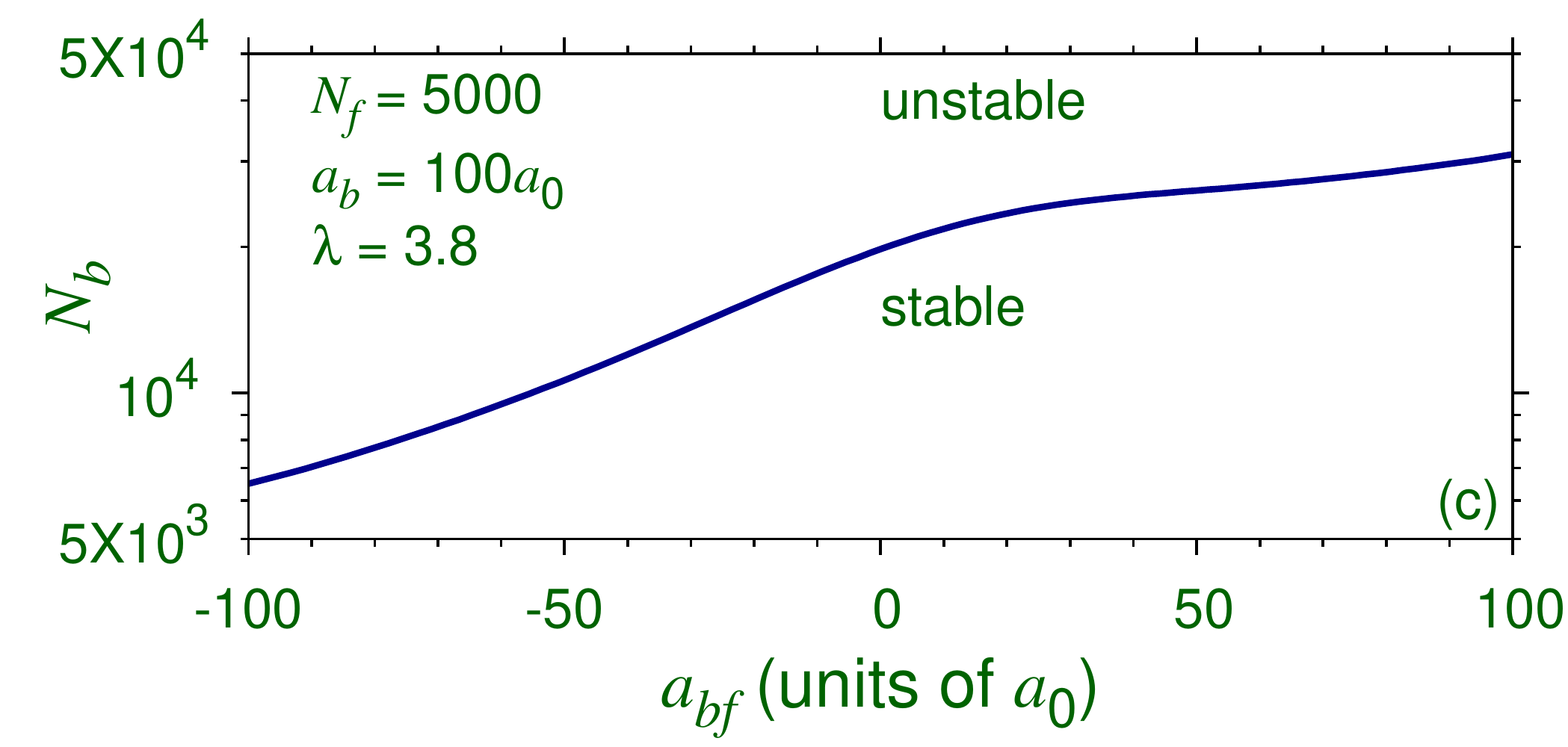}
\end{center}

\caption{(Color online) (a) Stability phase diagram of the  number of bosonic $^{164}$Dy atoms $N_b$ versus 
bosonic scattering length   $a_b$  for the binary $^{164}$Dy-$^{161}$Dy mixture for the number of 
fermionic $^{161}$Dy  atoms 
$N_f $= 0, 5000, and for the interspecies scattering length $a_{bf}$ = $110a_0$.
(b)  Stability phase diagram of $N_b$ versus $\lambda$
    for 
$N_f$ = 5000, $a_{bf}$ = $110a_0$, 
and  $a_b$ = $100a_0$.
(c) Stability phase diagram of $N_b$ versus $a_{bf}$
  for 
$N_f$ = 5000,   $a_b$ = $100a_0$.
Dipolar lengths $a_{dd}$ are taken as 
$a_{dd}^{(b)} = 132.7 a_0$, $a_{dd}^{(f)} = 130.3  a_0$, and $a_{dd}^{(bf)} = 131.5 a_0$ and oscillator length scale $l_0= 0.443$ $\mu$m.
}

\label{fig1}
\end{figure}

 \section{Numerical Results}

We perform  numerical 
calculation for the stability and dynamics of the binary dipolar $^{164}$Dy-$^{161}$Dy boson-fermion mixture using  realistic values of 
the parameters. 
We solve the dynamical equations (\ref{eq3}) and (\ref{eq4}) by the split-step Crank-Nicolson method \cite{CPC}{, using a space step of 0.1 $\sim$ 0.2 and time step of 0.001 $\sim$ 0.003.}
For the fermionic system of $^{161}$Dy atoms we take the trap frequencies 
$f_f = \{180,200,720\}$ Hz corresponding to the  geometrical mean angular frequency 
of $\omega_f= 2\pi \times 296$ Hz and trap aspect ratio
$\lambda_f =3.8$
 as used in the recent experiment \cite{122}. For the BEC of  $^{164}$Dy atoms we take 
the trap frequencies $f_b = \{195,205,760\}$ Hz 
 corresponding to the  geometrical 
mean angular frequency 
of $\omega_b= 2\pi \times 312$ Hz and trap aspect ratio
$\lambda_b =3.8$
as in the experiment \cite{boselev}.
The bosonic  oscillator length  $l_0=\sqrt{\hbar/m_b\omega_b}= 0.443$ $\mu$m and
 the fermionic oscillator length  $\sqrt{\hbar/m_f\omega_f}= 0.459$ $\mu$m.
The Dy atoms have a large magnetic dipole moment  $\mu = 10\mu_B$ with $\mu_B$ $(=9.27402 \times 10^{-24}$ m$^2$A) the Bohr magneton corresponding to the dipolar lengths
 $a_{dd}^{(b)}\equiv \mu_0\mu^2 m_b/(12\pi \hbar^2)\approx 132.7a_0$ for 
$^{164}$Dy, 
 $a_{dd}^{(f)}\equiv \mu_0\mu^2 m_f/(12\pi \hbar^2)\approx 130.3a_0$ for 
$^{161}$Dy, and 
 $a_{dd}^{(bf)}\equiv \mu_0\mu^2 m_R/(6\pi \hbar^2)\approx 131.5a_0$, with 
$a_0$ $ (=5.29\times 10^{-11}$ m) the Bohr radius, $\mu_0=4\pi\times 10^{-7}$ N/A$^2,$ $\hbar= 1.05457\times 10^{-34}$ m$^2$kg/s,1 amu = $1.66054\times 10^{-27}$ kg.   
Thus the dipolar interaction in Dy atoms is more that eight 
times larger than that in Cr atoms with a dipolar length $a_{dd}\approx 15a_0$ \cite{pfau}. 

  In the present study on the dynamics of collapse  the bosonic scattering length $a_b$ is taken as $100a_0$. 
First we study the stability of the binary dipolar boson-fermion mixture for a fixed total number of fermions $N_f$ = 5000
 and illustrate the results in Figs. \ref{fig1} through phase diagrams 
showing the critical number of bosons $N_b$ in a stable 
boson-fermion mixture versus (a) the bosonic scattering length 
$a_b$, (b) trap aspect ratio $\lambda=\lambda_b=\lambda_f$, 
and (c) the boson-fermion scattering length $a_{bf}$ 
keeping other variables fixed at constant values, e.g., (a) 
$a_{bf}$ = $110a_0$, $\lambda =3.8$, (b) 
$a_{bf}$ = $110a_0$, $a_b$ = $100a_0$, and 
(c) $a_b$ = $100a_0$, and $\lambda =3.8$.
Because of the 
strong interspecies and intraspecies dipolar interactions, the binary 
mixture becomes unstable beyond a total number of atoms, independent of 
the other parameters, as shown in Figs. \ref{fig1}. 
{ To identify the region of instability of the binary system, time 
evolution with  
dynamical equations (\ref{gp3db}) and (\ref{gp3df}) is carried to very large time (about 50 to 100 units of time). The system is considered stable if this 
procedure leads to finite converged densities.}
Similar instability 
for larger net dipolar interaction was noted in case of both single-component 
\cite{cola,pfau}
and binary \cite{dipolarbin} dipolar BEC.
The increased 
repulsion for larger intraspecies and interspecies contact interactions 
appearing for large values of $a_b$ or 
$a_{bf}$, respectively, favors stability, and hence 
can accommodate a larger number of $^{164}$Dy atoms as can be seen in 
Figs. \ref{fig1} (a) and (c). For fixed values of contact interactions, 
a disk shape favors stability as can be seen in Fig. \ref{fig1} (b).
The phase diagram for $N_f=0$ in Fig. \ref{fig1} (a) reveals that 
the extra dipolar interaction in the boson-fermion mixture makes the system 
more vulnerable to collapse for repulsive contact interactions, when compared with the pure bosonic system.

It is known that, unlike the nondipolar degenerate fermion gas, 
the polarized dipolar fermion gas collapses for increased dipolar interaction \cite{pucol}. But the 
tendency to collapse is much enhanced in the dipolar boson-fermion mixture in the presence 
of attractive interspecies interaction.  
Hence,
to study the collapse dynamics in the boson-fermion mixture
we introduce instability by suddenly changing 
the interspecies contact interaction from repulsive to attractive which can be achieved by varying a background magnetic field near a Feshbach resonance \cite{fesh}. A similar variation of the interspecies contact interaction from 
repulsive to attractive
has been demonstrated to lead to a net attraction in an otherwise repulsive binary mixture resulting in soliton formation  \cite{binaatt2} and collapse
 \cite{binaatt1}. 
During collapse
we maintain all the dipolar interactions and the intraspecies bosonic interaction at the respective initial values. From Fig. \ref{fig1} (c) we find that a larger number of about 30000 $^{164}$Dy atoms  can be accommodated for $a_{bf}$ = $100a_0$ than about 
6000 $^{164}$Dy
atoms 
for $a_{bf}$ = $-100a_0$.    Hence for the study of collapse 
dynamics we consider an initial mixture of 20000 $^{164}$Dy atoms and 5000 
$^{161}$Dy atoms and initiate collapse by suddenly jumping the interspecies scattering length 
$a_{bf}$ from $100a_0$ to $-100a_0$. 
From Fig. \ref{fig1} (c) we find that at  $a_{bf}$ = $ - 100a_0$  the binary dipolar mixture of 20000 $^{164}$Dy atoms and 5000 
$^{161}$Dy atoms becomes unstable to collapse. 
It is pertinent to mention that the same binary
boson-fermion system with all dipolar interactions set to
zero is absolutely stable and does not collapse even for the
attractive interspecies scattering length $a_{bf}$
=  $ - 100a_0$. Hence the collapse studied here is caused solely by
the dipolar interaction. This is further substantiated by
the higher-partial-wave shape in density distribution in
Fig.  \ref{fig5}. The numerical simulation was carried out in a
sufficiently large 3D box of size $38\times38\times 25$ and there was
no reflection of matter wave from the boundary.

For the  study of the collapse dynamics we have to fix the values of three-body 
loss rates $k_3^{(bb)}$ and $k_3^{(bf)}$ in Eqs. (\ref{eq3}) and (\ref{eq4}). We did the calculation for three sets of dimensionless  loss rates: (i) $k_3^{(bb)}= 0.00005$ and $k_3^{(bf)}=0.0001$, (ii) $k_3^{(bb)}= 0.0005$ and $k_3^{(bf)}=0.001$, and 
 (iii) $k_3^{(bb)}= 0.00001$ and $k_3^{(bf)}=0.00004$.
 These correspond to the
following physical loss rates (i) $K_3^{(bb)}\equiv k_3^{(bb)}\omega_b l_0^6=7.5\times 10^{-28}$  cm$^6$/s and 
$K_3^{(bf)}\equiv  k_3^{(bf)}\omega_b l_0^6= 1.5\times 10^{-27}$  cm$^6$/s;
(ii) $K_3^{(bb)}=7.5 \times 10^{-27}$  cm$^6$/s and
 $K_3^{(bf)}= 1.5\times 10^{-26}$  cm$^6$/s; and 
 (iii) $K_3^{(bb)}=1.5 \times 10^{-28}$  cm$^6$/s and
 $K_3^{(bf)}= 6\times 10^{-28}$  cm$^6$/s;
where we used $\omega_b=\omega=2\pi \times 312$ Hz and $l_0= 0.443$ $\mu$m. In the absence of accurate experimental numbers for these loss rates, we use these sets of values which seem to be quite realistic and comparable to the known experimental rates for other atoms. For example, for dipolar $^{52}$Cr atoms the loss rate  
$K_3= 2\times 10^{-28}$  cm$^6$/s  was used \cite{12},
in a $^{87}$Rb-$^{40}$K boson-fermion mixture a loss rate of $K_3=  2\times 10^{-27}$  cm$^6$/s was measured \cite{modugno}, 
and for $^{85}$Rb atoms the experimental loss rate was $K_3\approx  5\times 10^{-25}$  cm$^6$/s \cite{lossrate}.  As the collapse dynamics of the binary 
mixture 
could be sensitive to loss rate, the three rates (i), (ii), and (iii)
chosen here correspond to medium, strong, and mild loss rates, respectively. 
The collapse dynamics is expected to be distinct as the loss rates are changed from mild to strong and this fact motivates the study with three rates above.      
The third  set (iii) with the smallest loss rates 
produces a slower atom loss in the beginning and larger final residual remnant  
states after the collapse. The second set (ii) with the largest loss rates leads to a faster atom loss in the beginning and smaller final residual remnant  
states.
The above sets of loss rates lead to a sizable amount of atom loss in less than 10 units of time as in the collapse of a nondipolar BEC of $^{85}$Rb atoms \cite{donley}.

\begin{figure}
\begin{center}
\includegraphics[width=\linewidth]{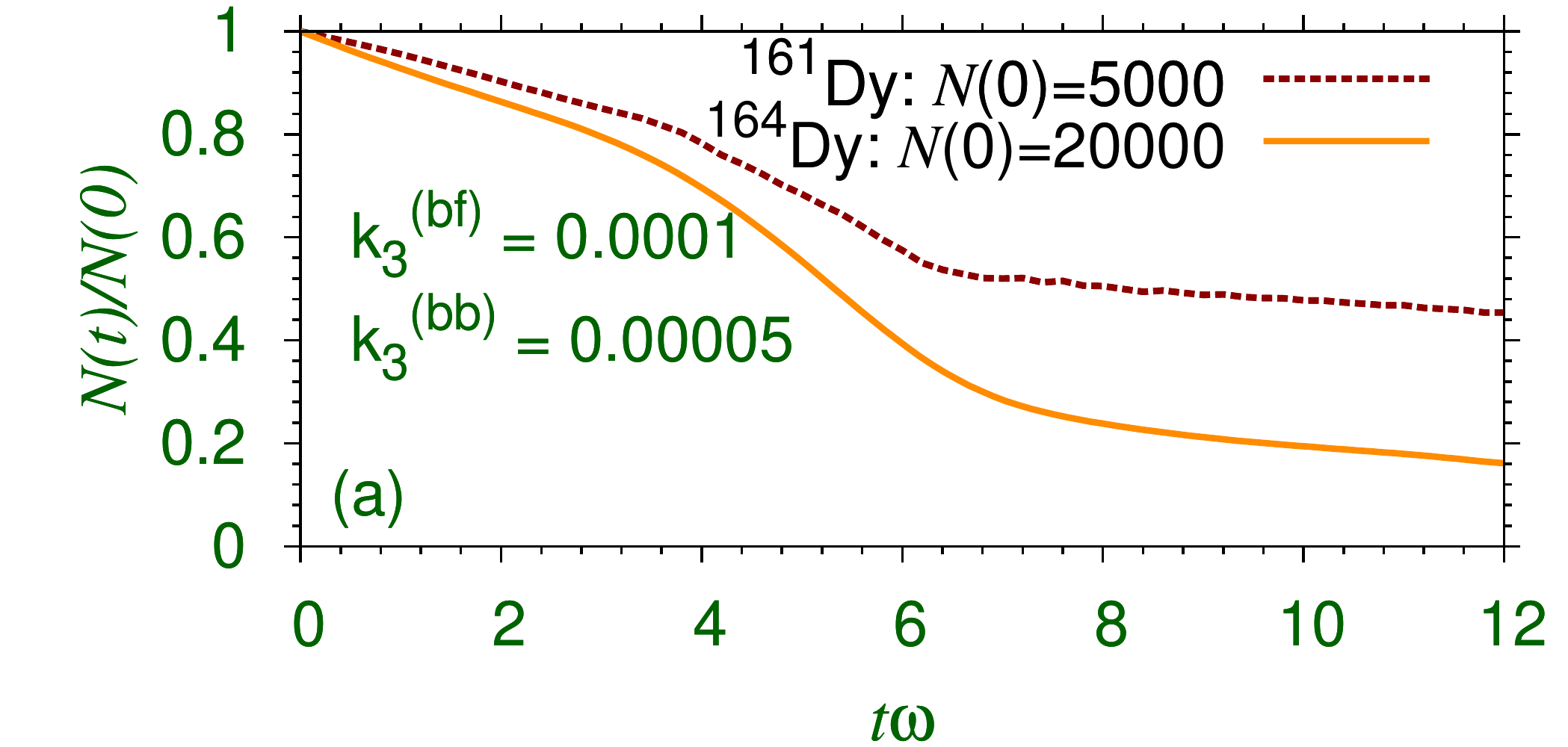}
\includegraphics[width=.45\linewidth]{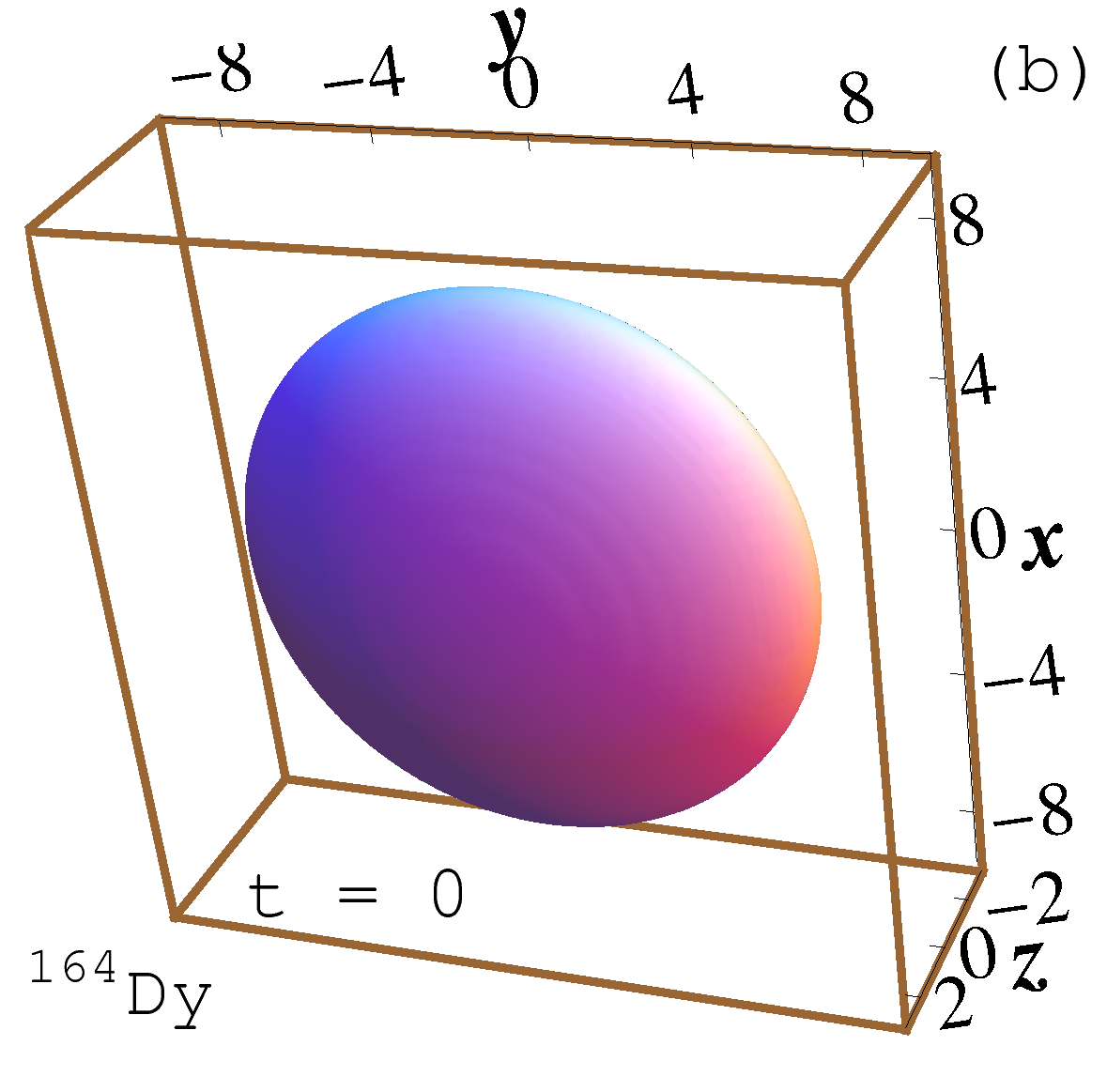}
\includegraphics[width=.45\linewidth]{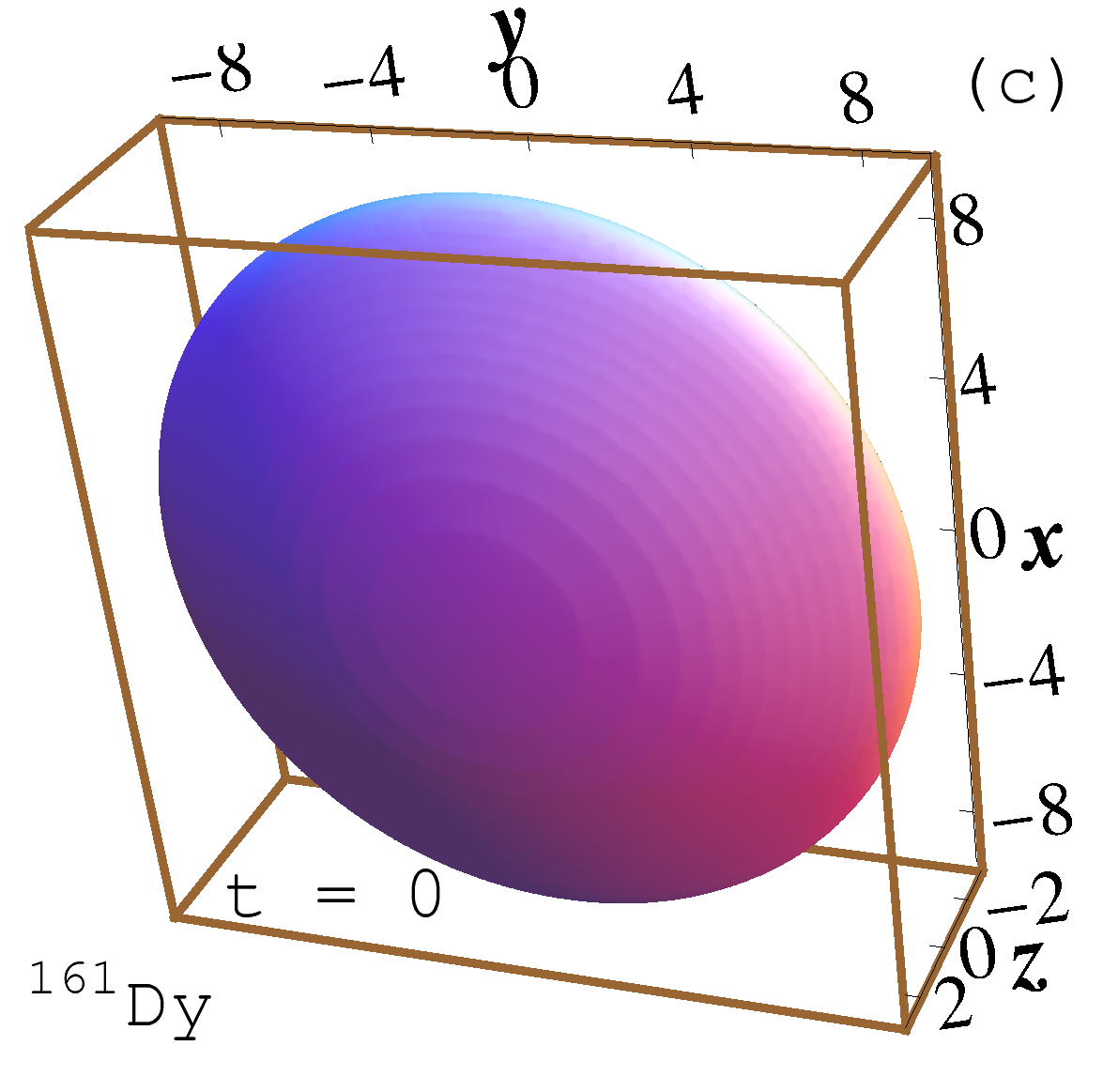}
\includegraphics[width=.45\linewidth]{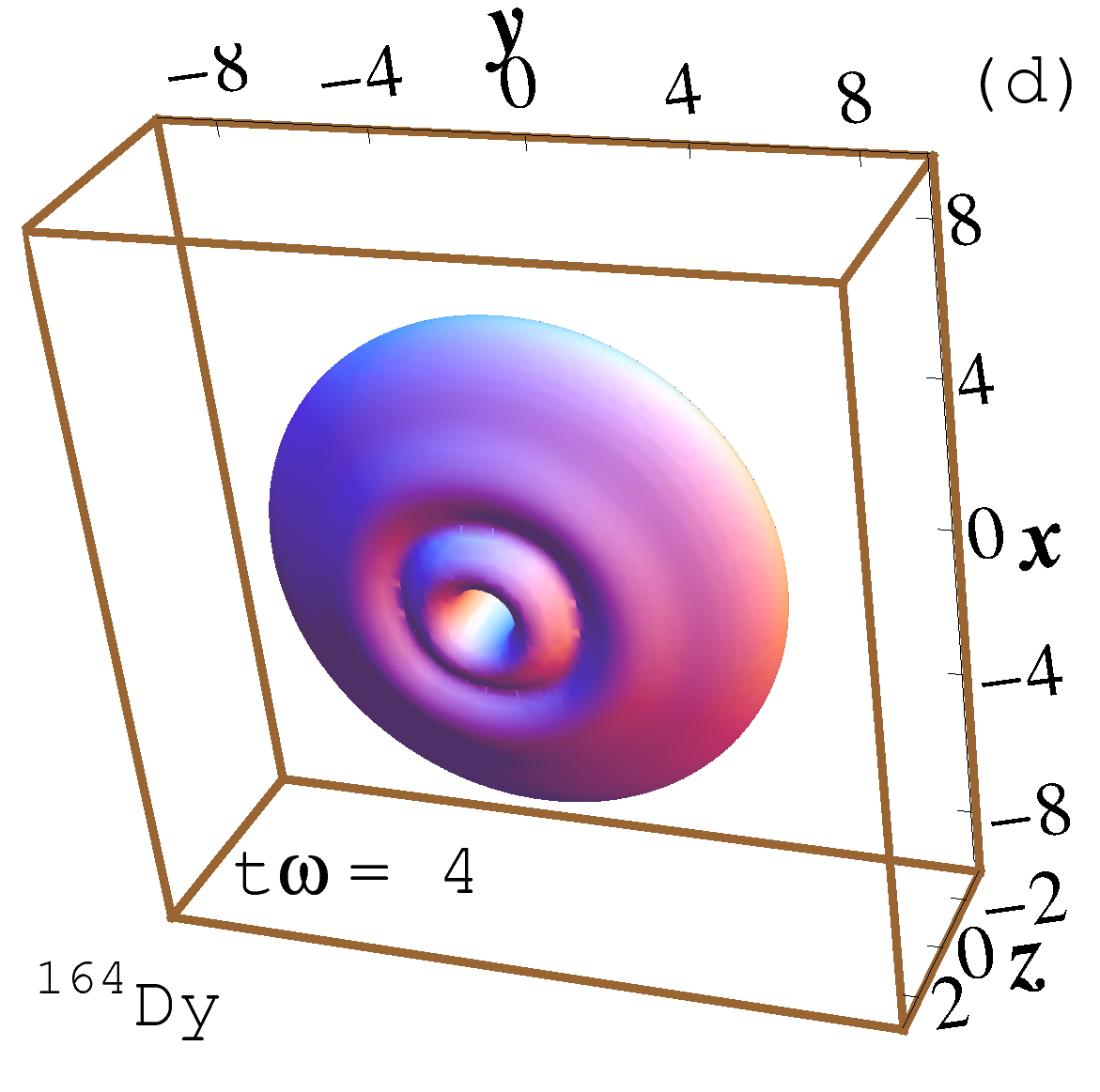}
\includegraphics[width=.45\linewidth]{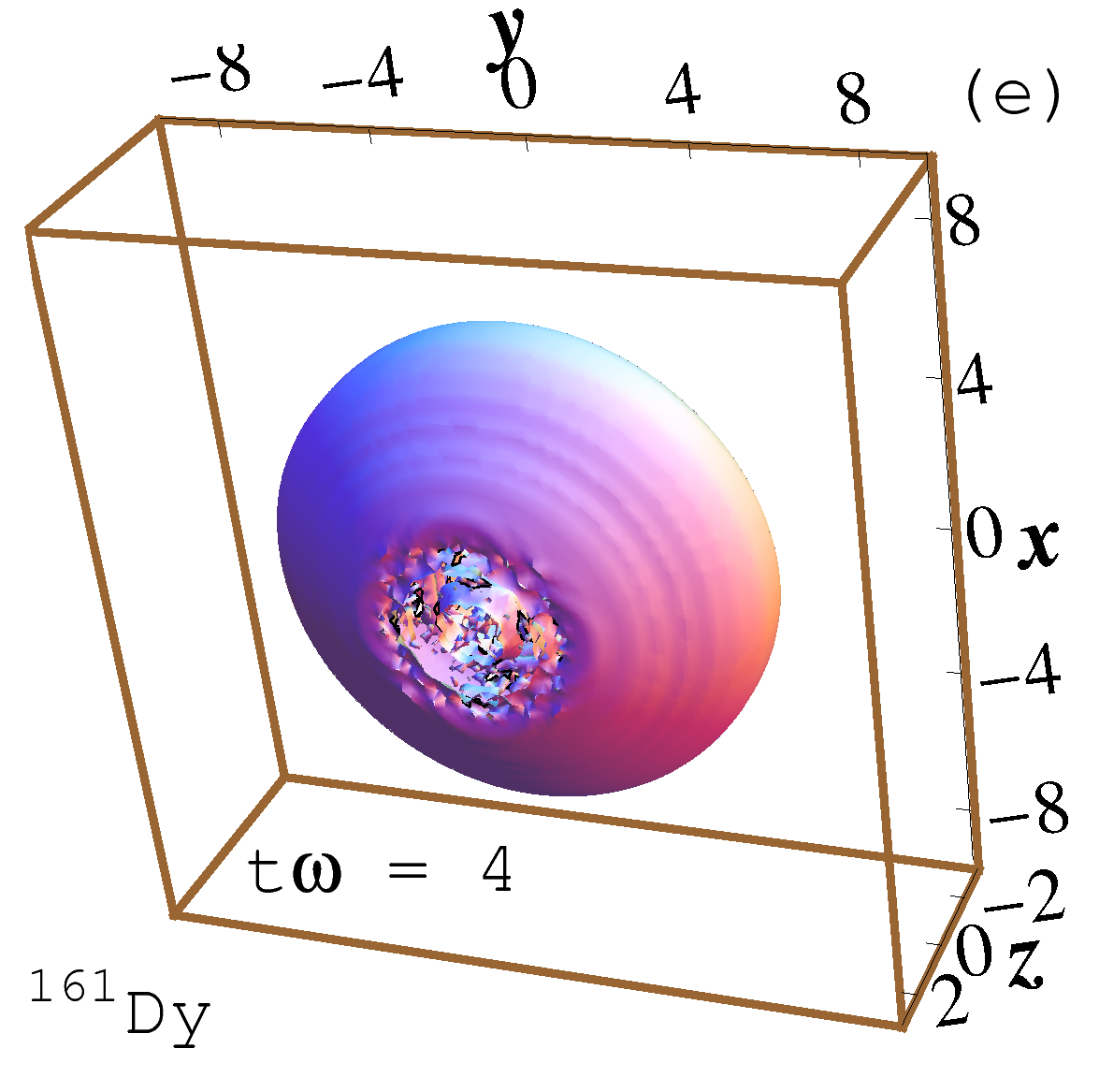}
\includegraphics[width=.45\linewidth]{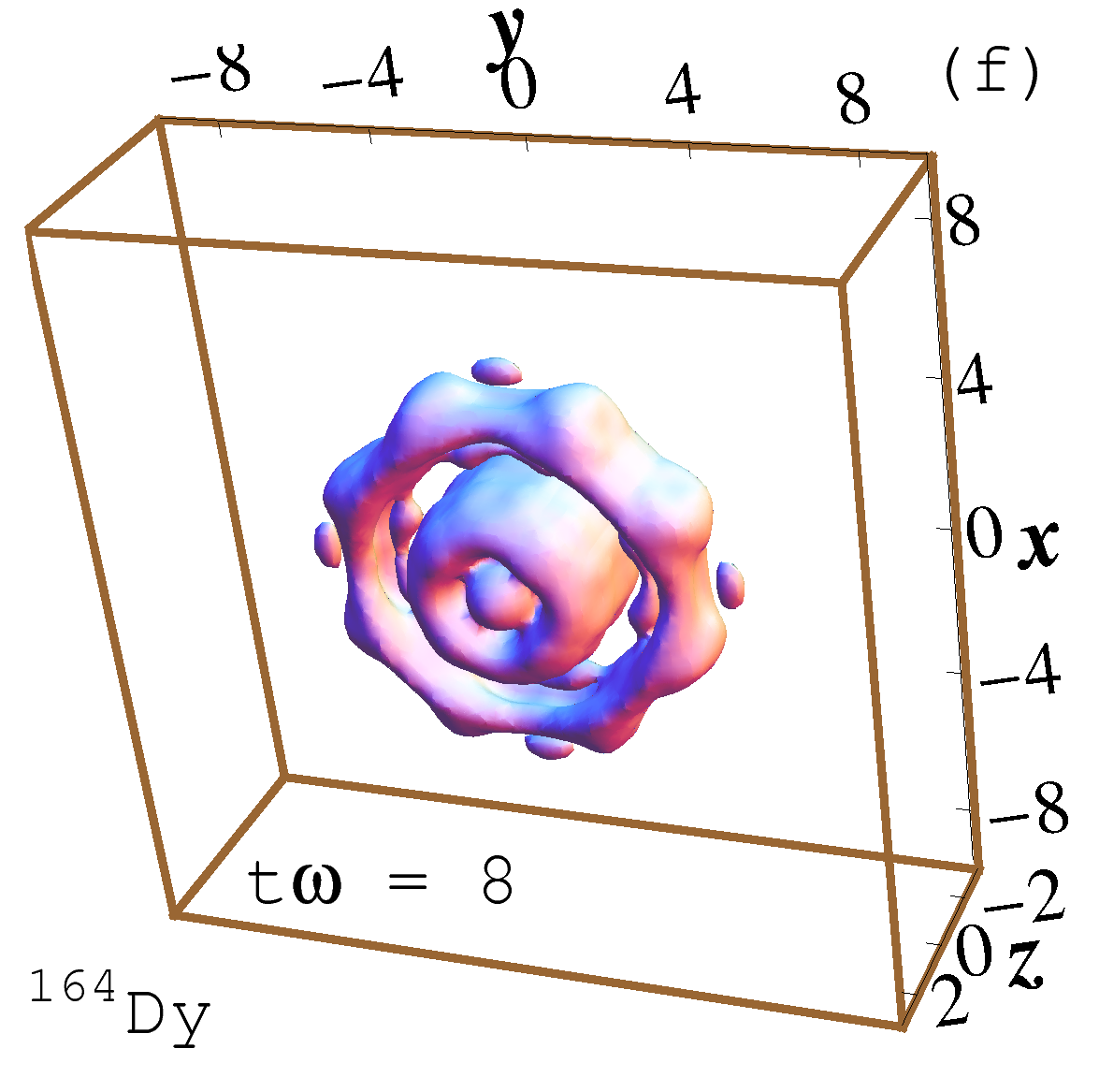}
\includegraphics[width=.45\linewidth]{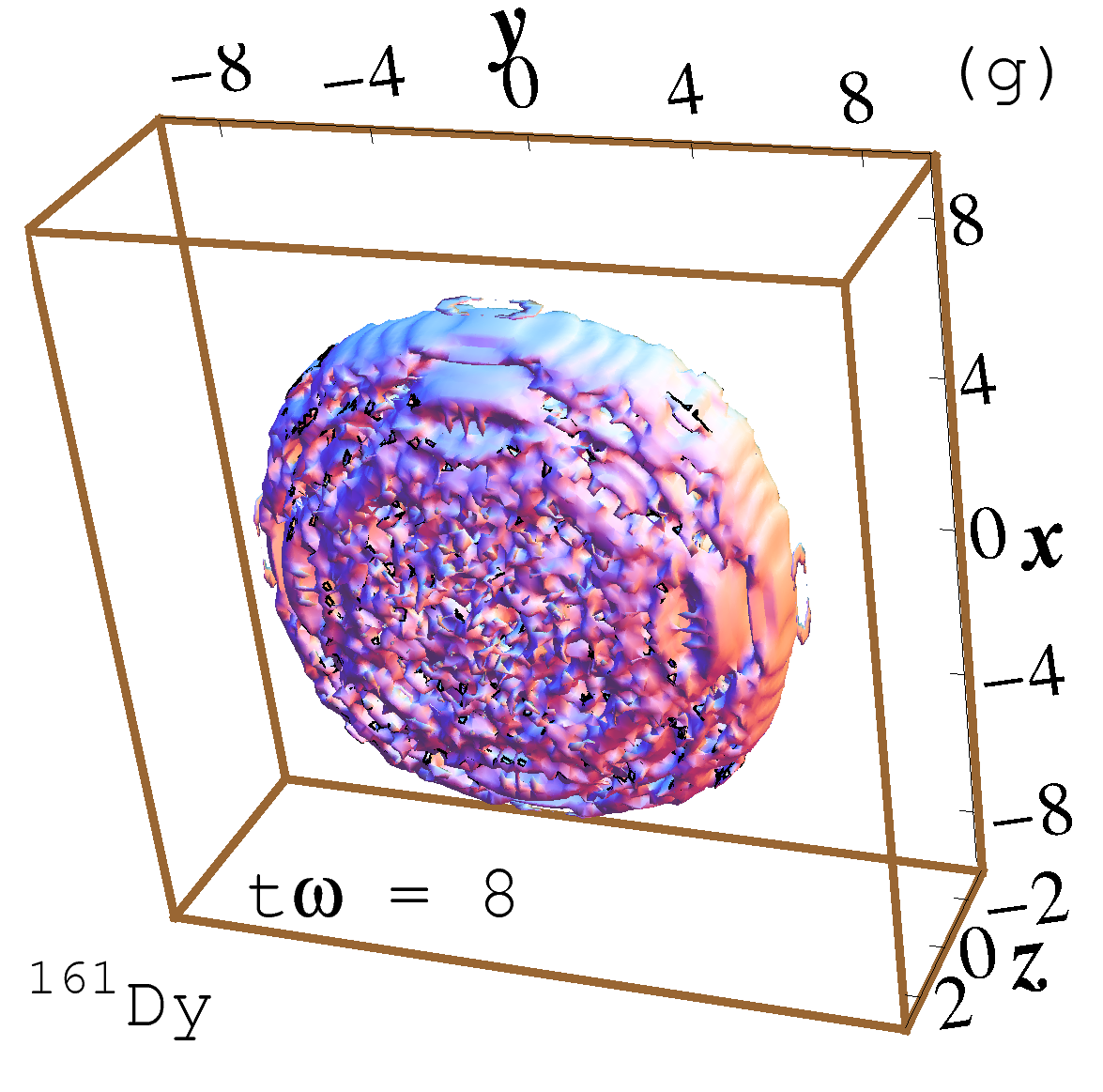}
\includegraphics[width=.45\linewidth]{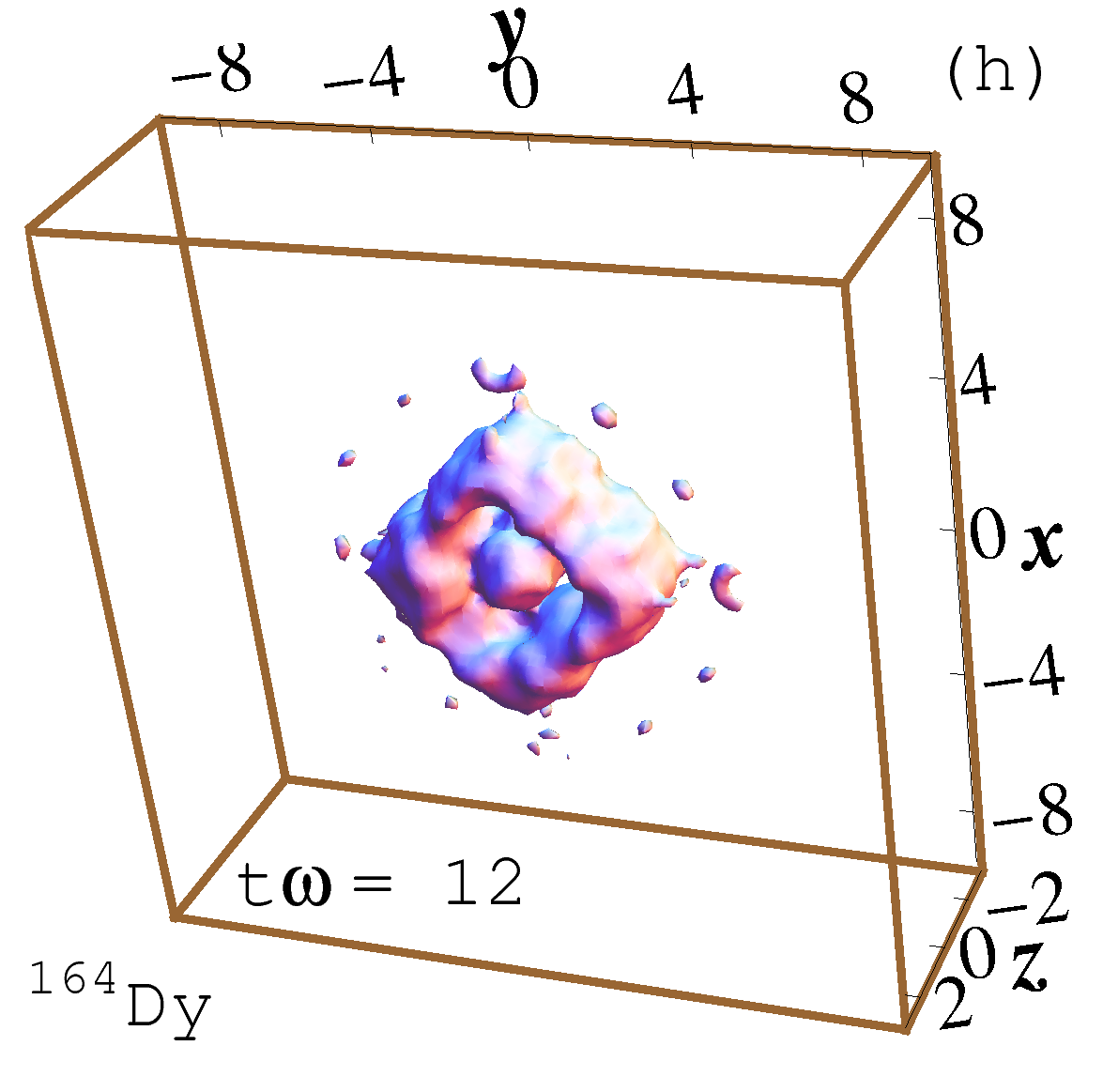}
\includegraphics[width=.45\linewidth]{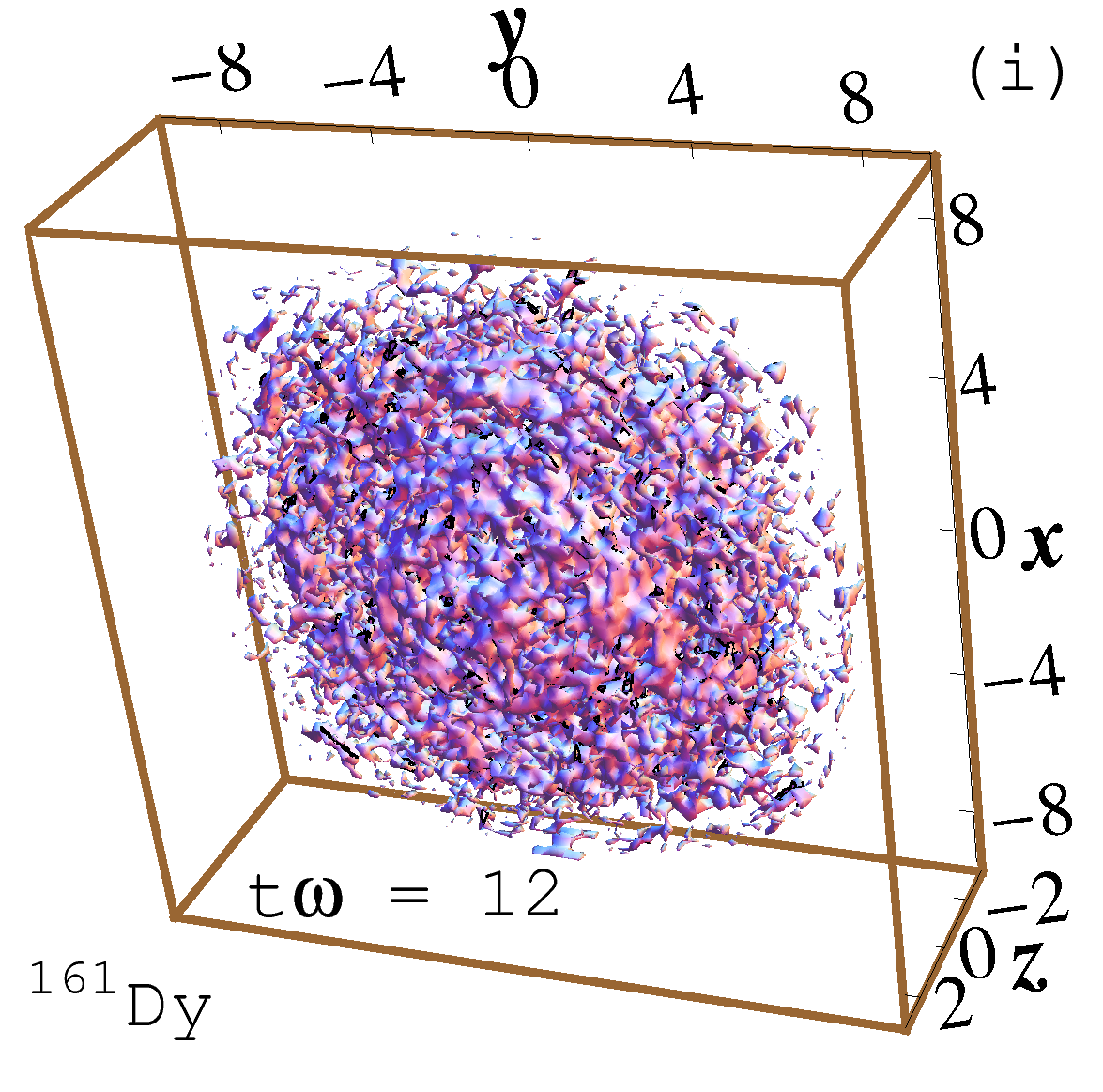}
\end{center}

\caption{(Color online) (a) Relative number of atoms $N(t)/N(0)$ versus time 
during collapse of a binary mixture of 20000 
$^{164}$Dy  and 5000 
$^{161}$Dy atoms
initiated by jumping $a_{bf}$ from 
$100a_0$ to $-100a_0$.  Isodensity profile of $^{164}$Dy and $^{161}$Dy atoms
at (dimensionless) times $t\omega=0,4,8,12$ are shown in (b) $-$ (h). Parameters used 
$a_{dd}^{(b)} = 132.7 a_0$, $a_{dd}^{(f)} = 130.3  a_0$,
 $a_{dd}^{(bf)} = 131.5 a_0, a_{b}=100a_0, \lambda=3.8,$ 
 $K_3^{(bb)}=7.5\times 10^{-28}$ cm$^6$/s,
$K_3^{(bf)}=1.5\times 10^{-27}$ cm$^6$/s,
density on contour 0.0005, and $\omega\equiv \omega_b$. Lengths and densities are in units of 
oscillator length $l_0$ and $l_0^{-3}$. }

\label{fig2}
\end{figure}

\begin{figure}
\begin{center}
\includegraphics[width=\linewidth]{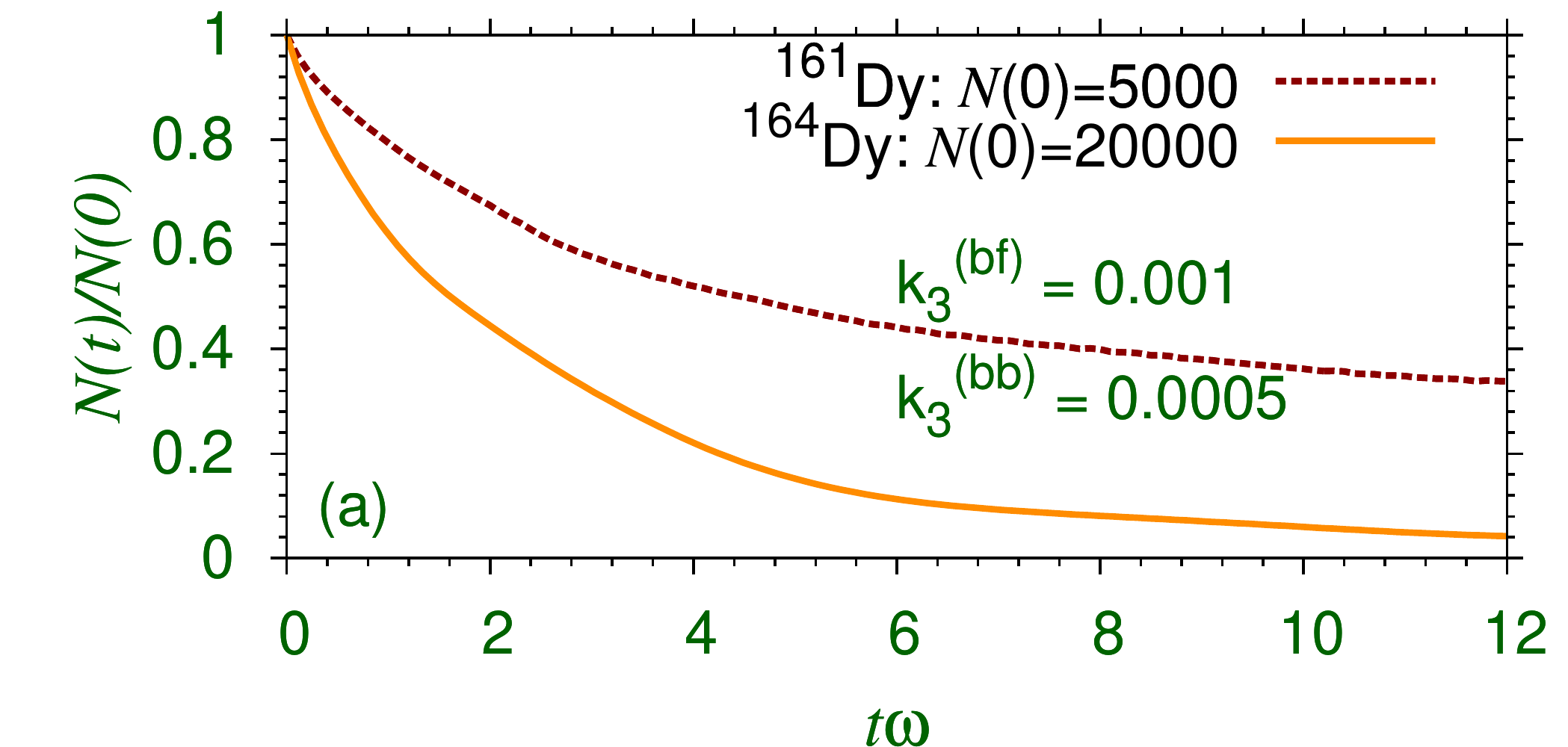}
\includegraphics[width=.49\linewidth]{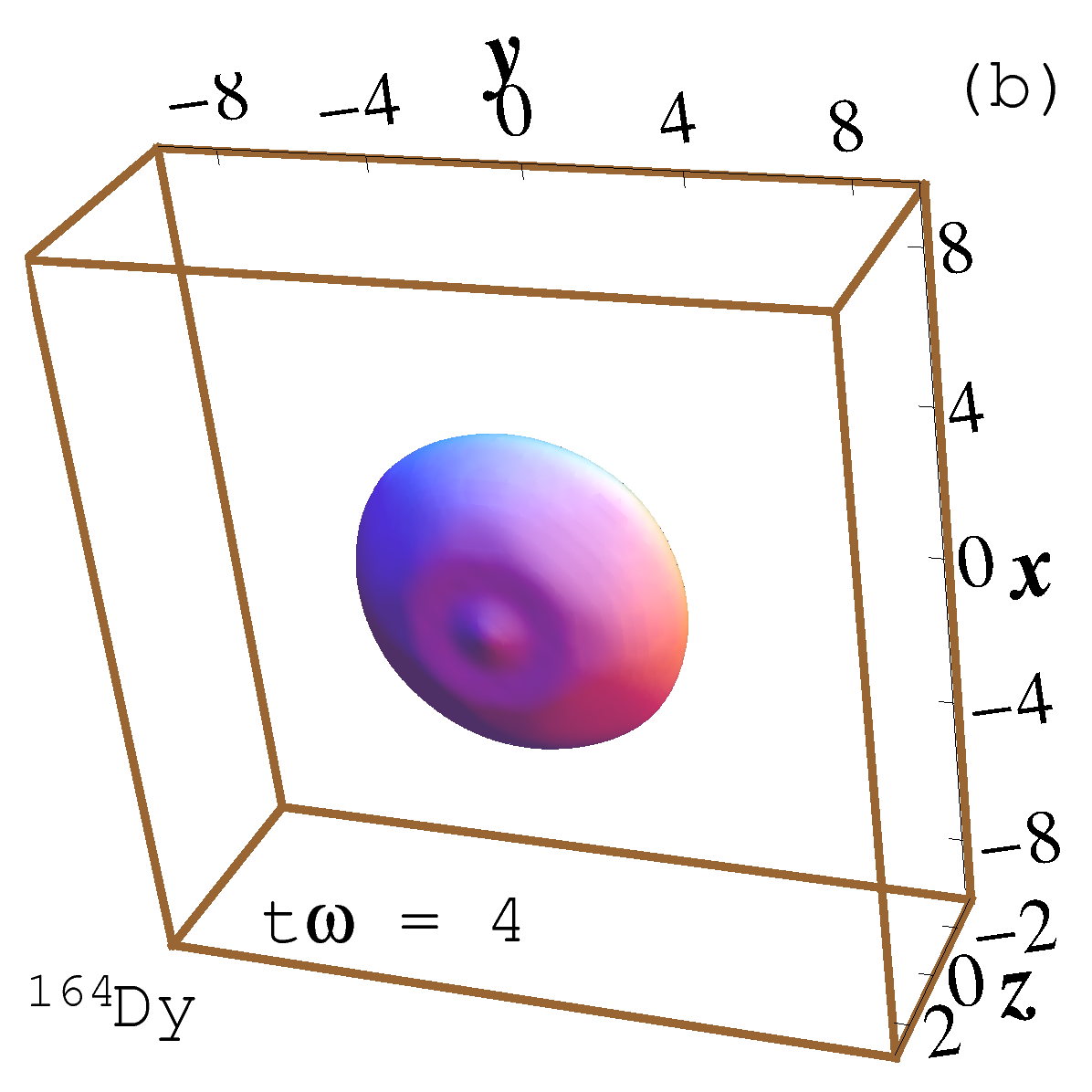}
\includegraphics[width=.49\linewidth]{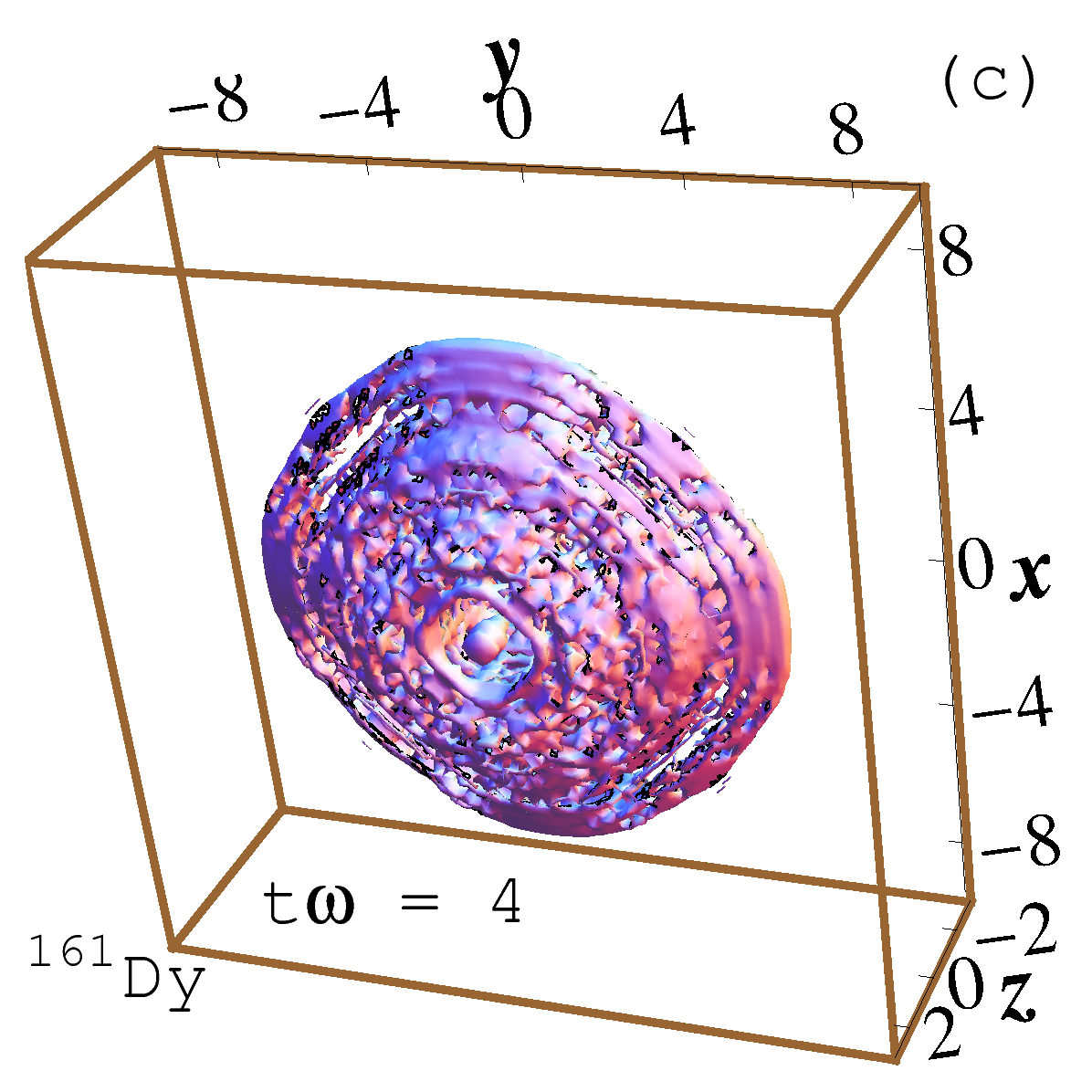}
\includegraphics[width=.49\linewidth]{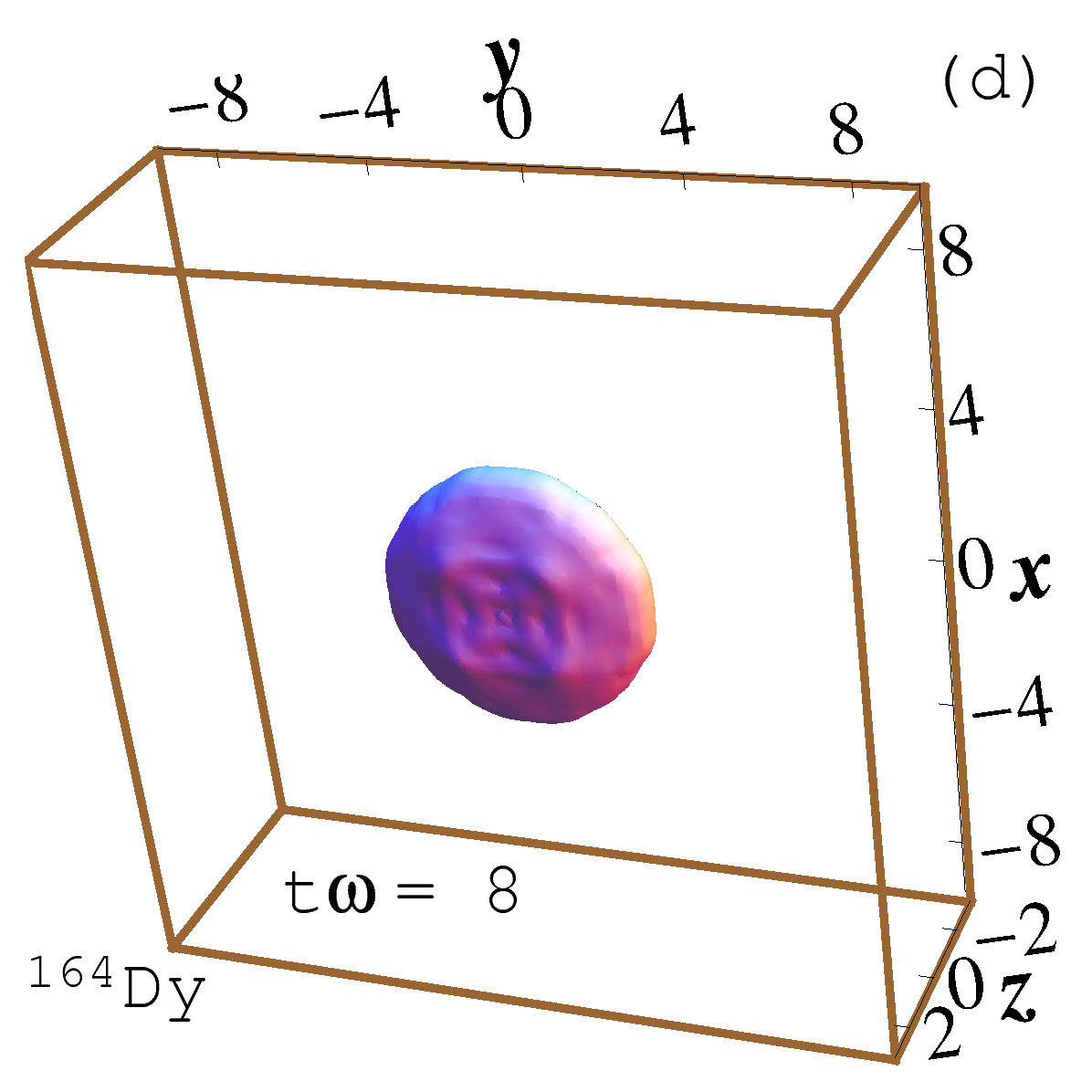}
\includegraphics[width=.49\linewidth]{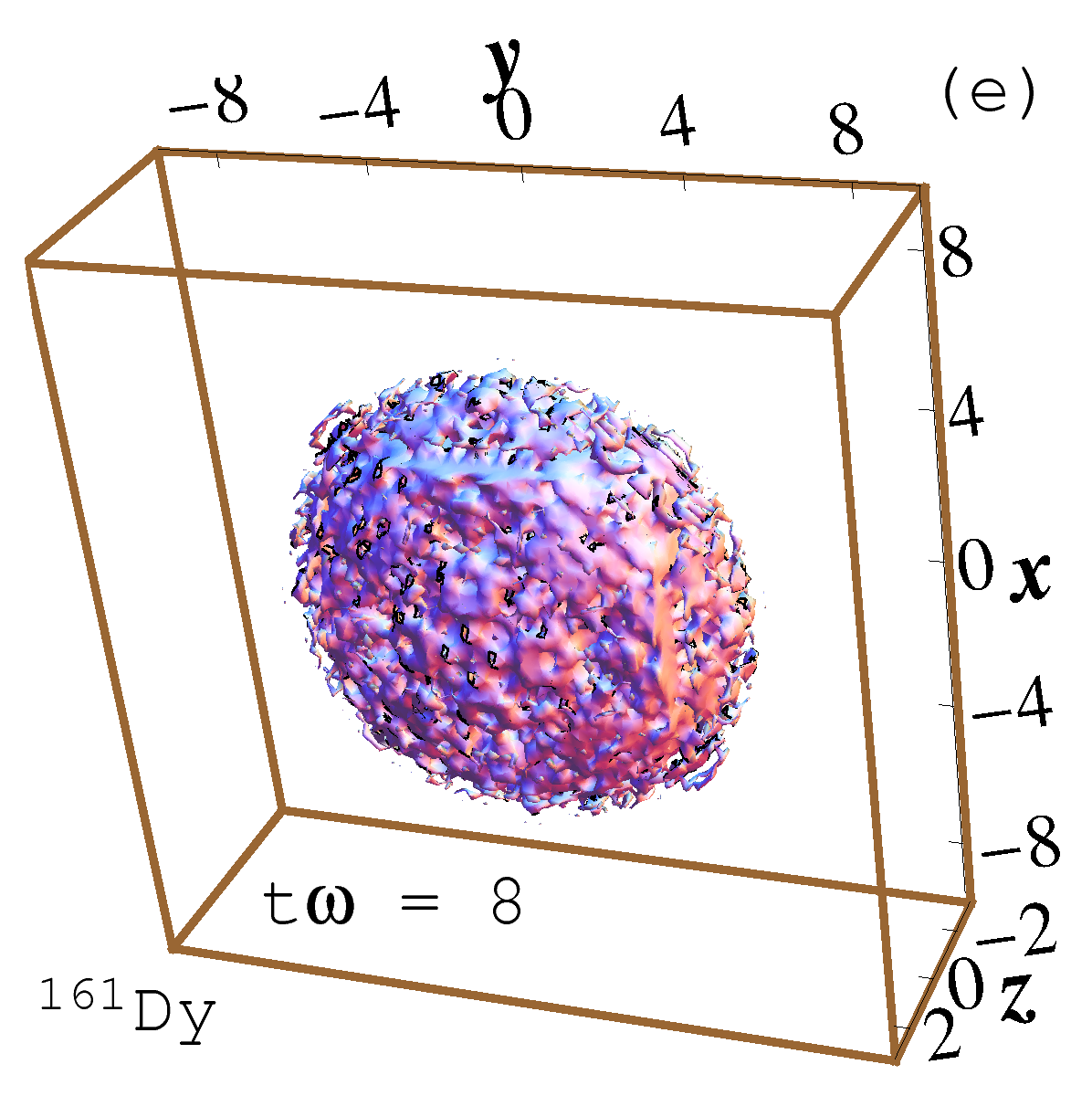}
\end{center}

\caption{(Color online)(a) Same as in Fig. \ref{fig2} (a) for loss rates 
$K_3^{(bb)}=7.5\times 10^{-27}$ cm$^6$/s,
$K_3^{(bf)}=1.5\times 10^{-26}$ cm$^6$/s.
Isodensity profile of $^{164}$Dy and $^{161}$Dy atoms
at times $t\omega=4,8$ are shown in (b) $-$ (e) for these loss rates. Other parameters are the 
same as in Fig. \ref{fig2}.}

\label{fig3}
\end{figure}

\begin{figure}
\begin{center}
\includegraphics[width=\linewidth]{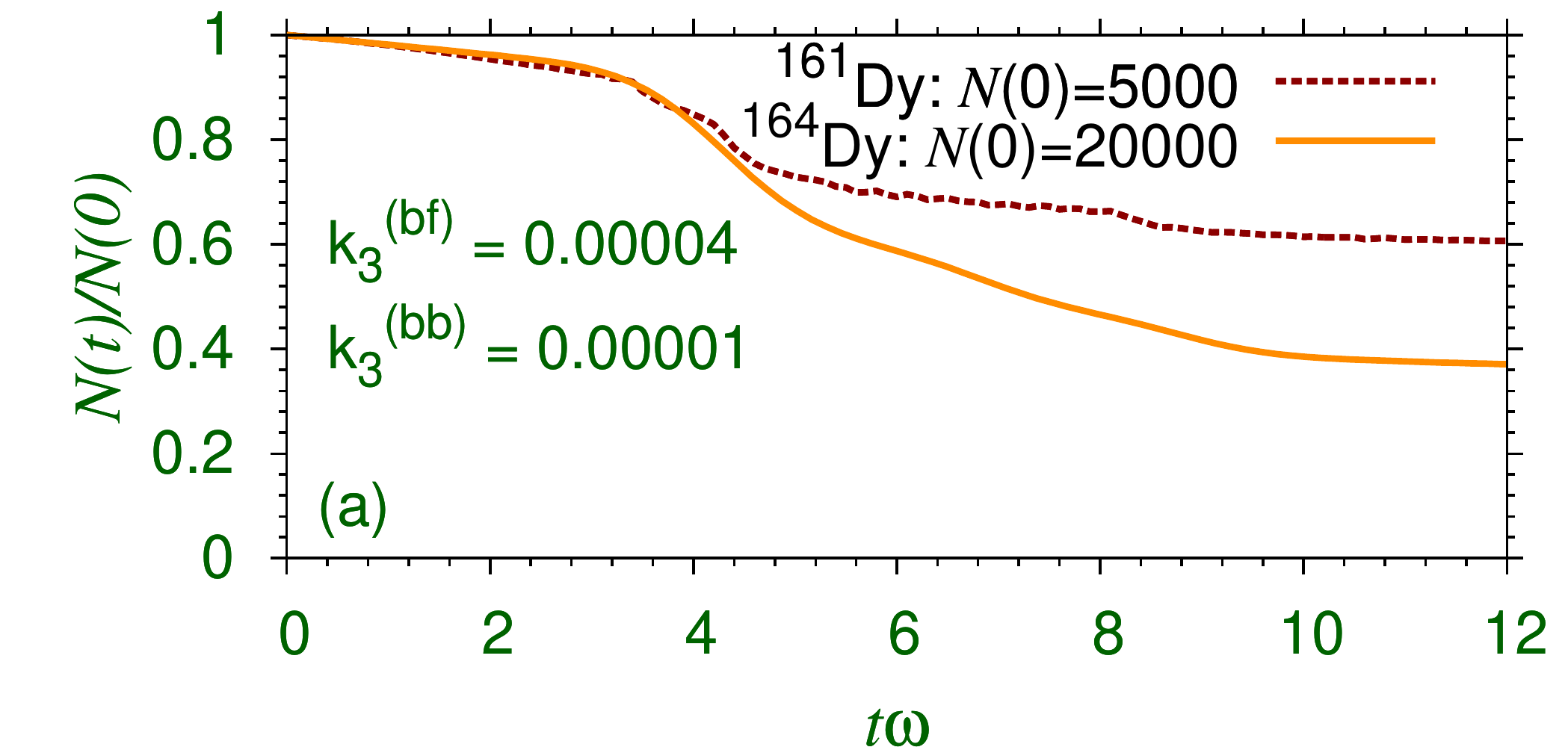}
\includegraphics[width=.49\linewidth]{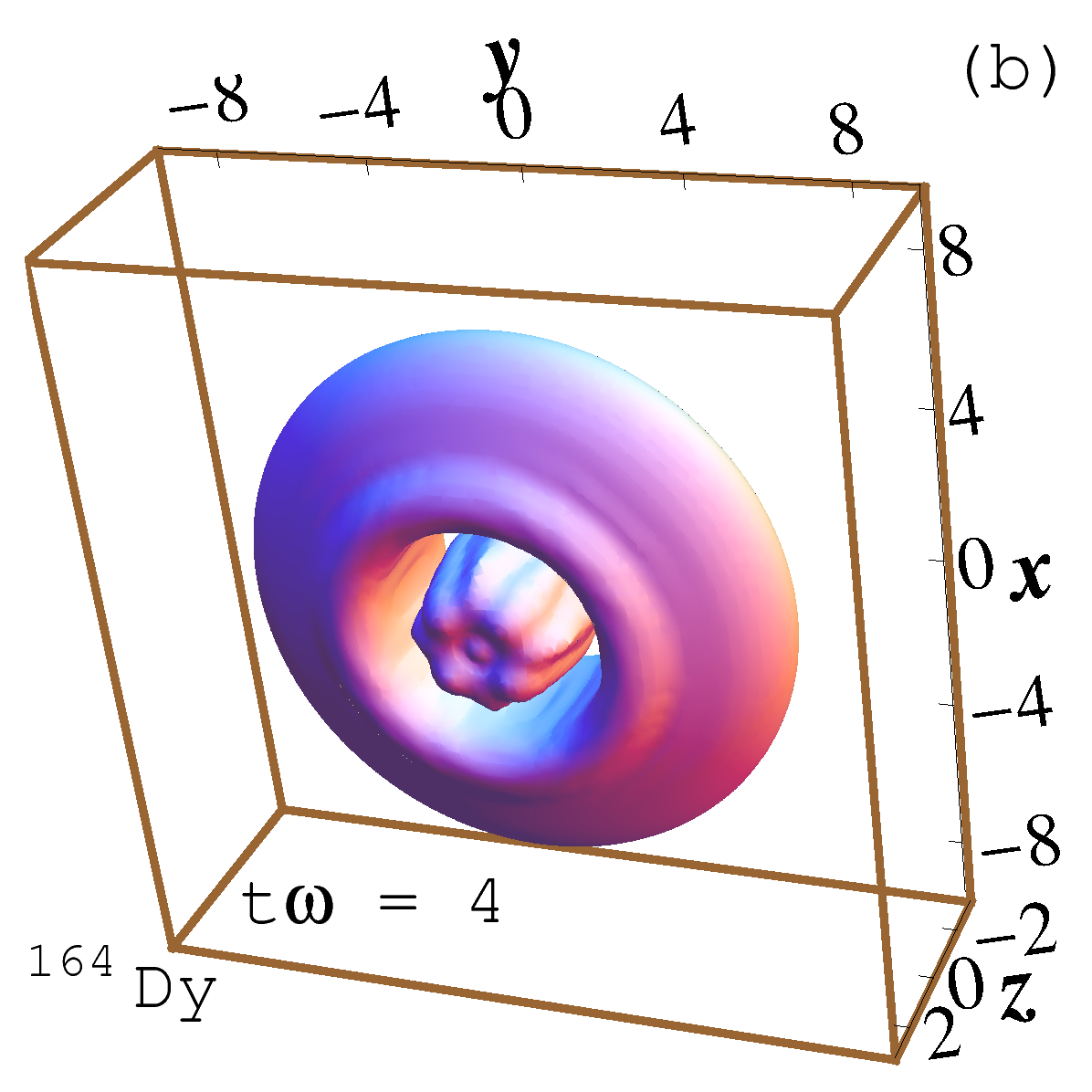}
\includegraphics[width=.49\linewidth]{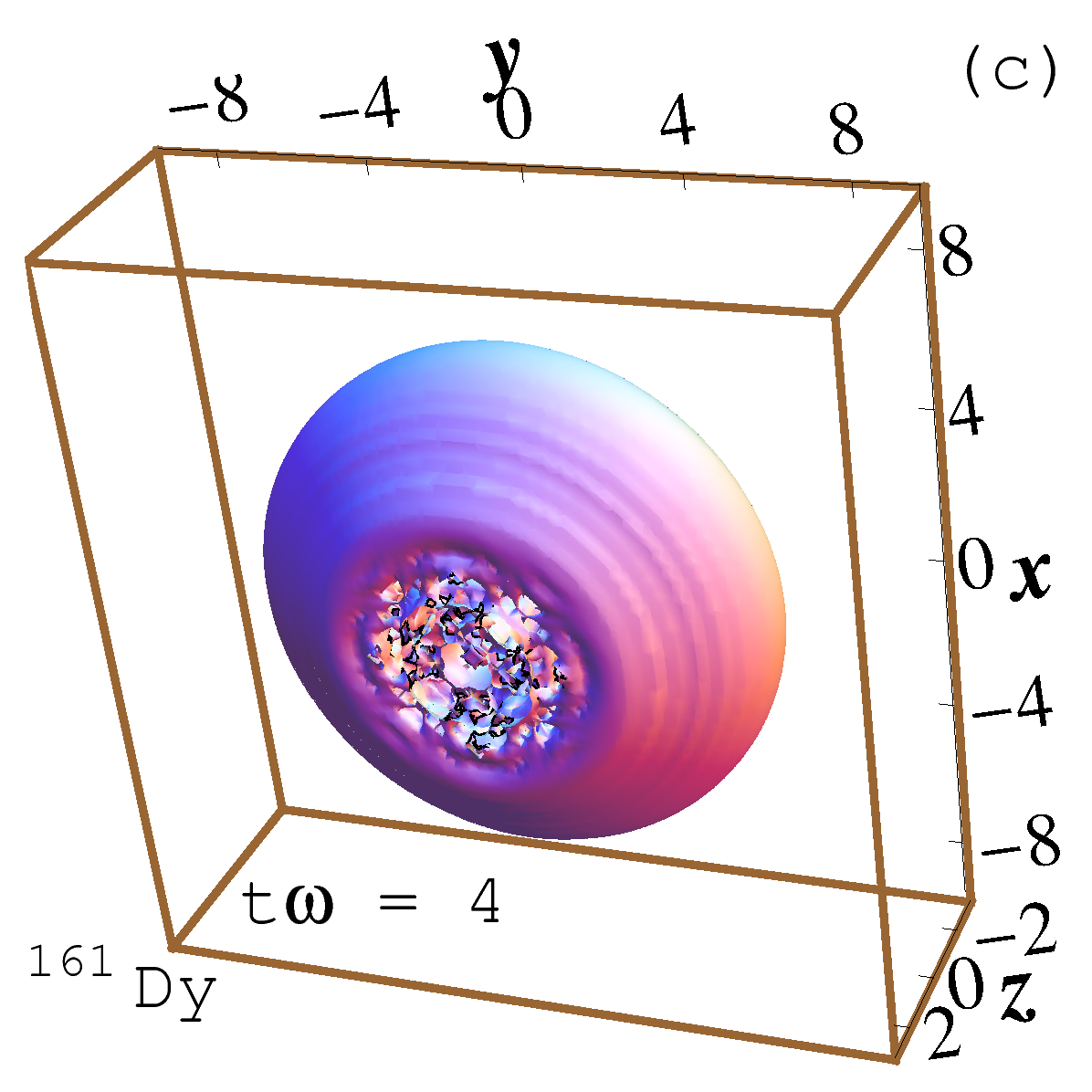}
\includegraphics[width=.49\linewidth]{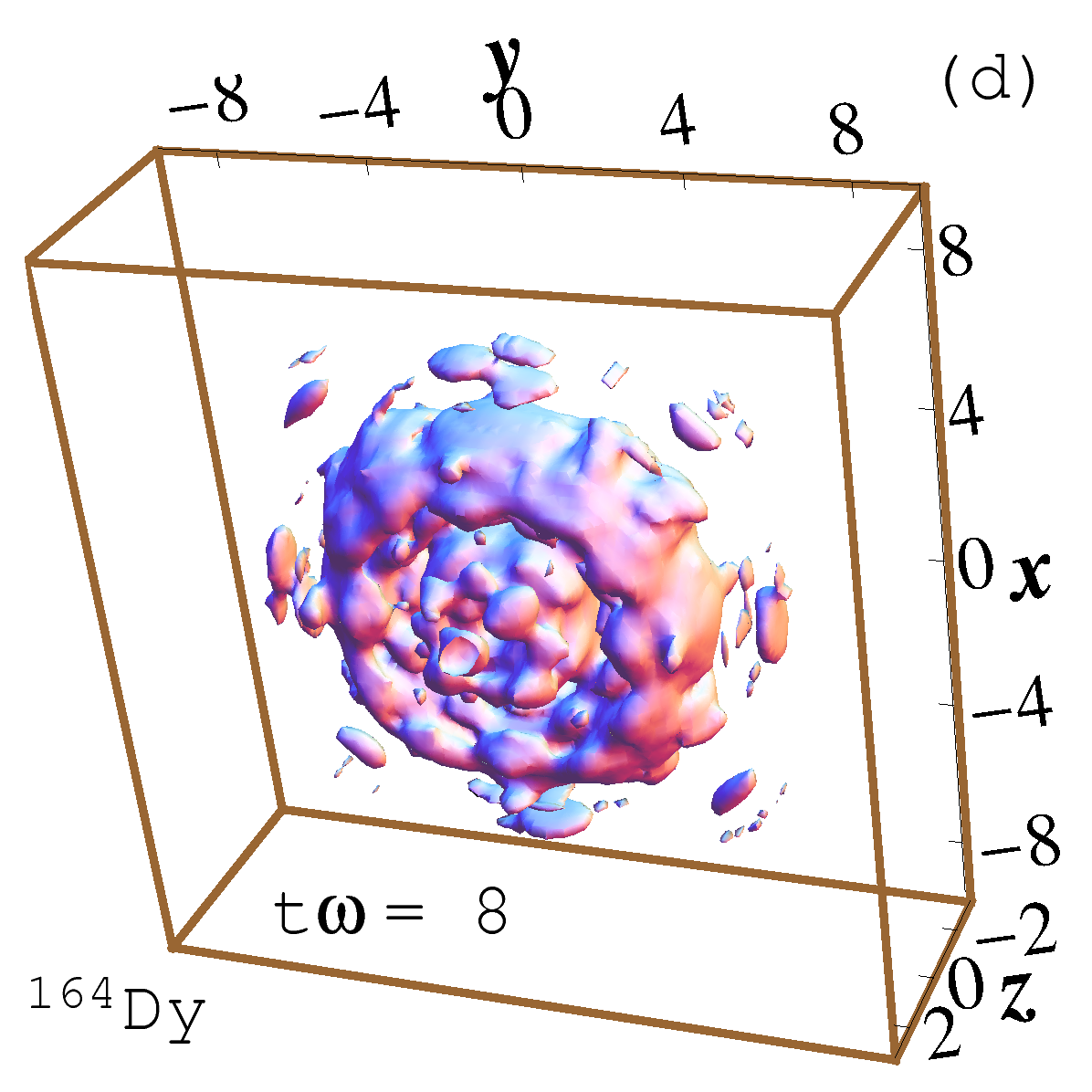}
\includegraphics[width=.49\linewidth]{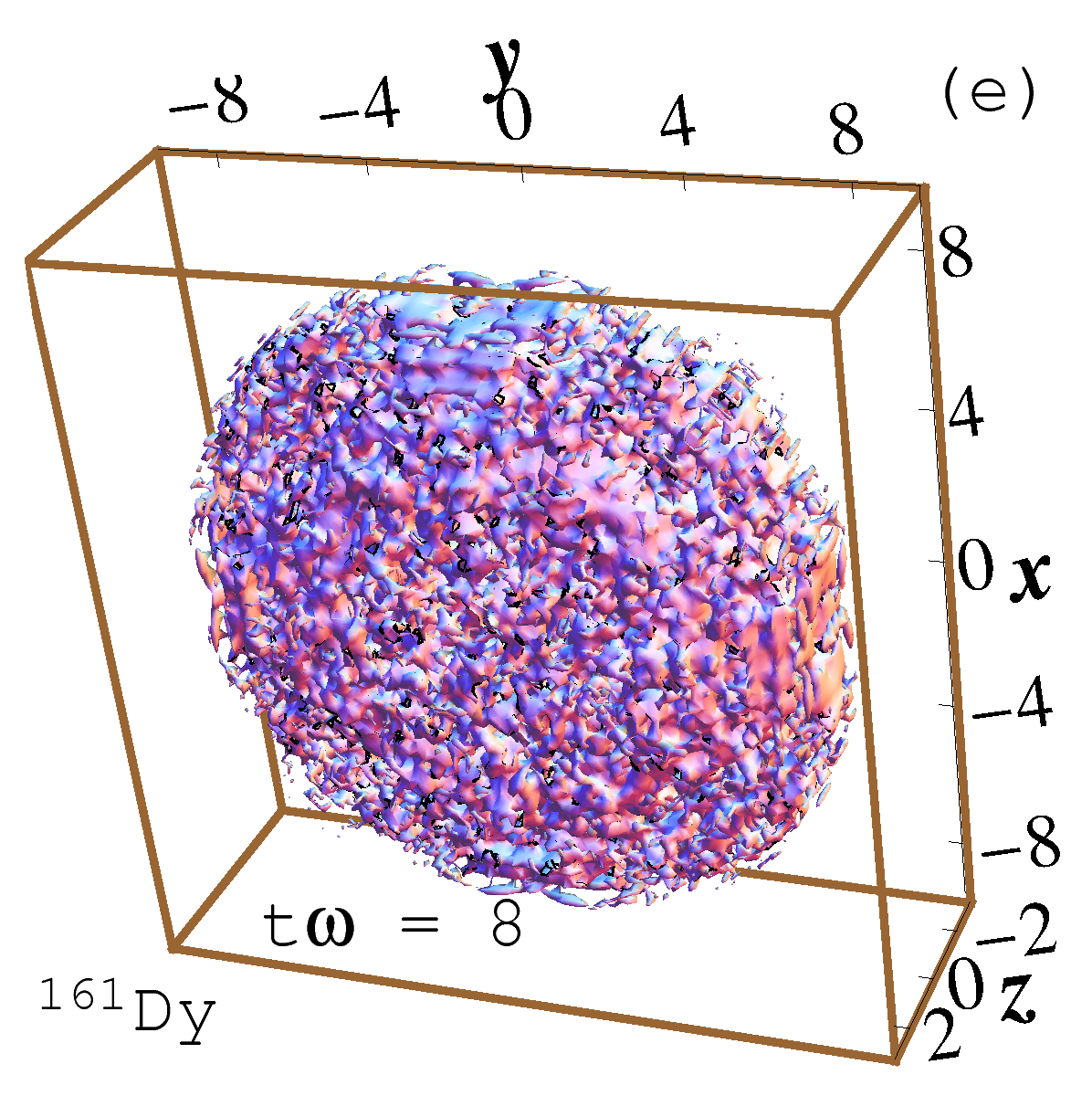}
\end{center}

\caption{(Color online)(a) Same as in Fig. \ref{fig2} (a) for loss rates 
 $K_3^{(bb)}=1.5\times 10^{-28}$ cm$^6$/s,
$K_3^{(bf)}=6\times 10^{-28}$ cm$^6$/s,
Isodensity profile of $^{164}$Dy and $^{161}$Dy atoms
at times $t\omega=4,8$ are shown in (b) $-$ (e) for these loss rates. Other parameters are the 
same as in Fig. \ref{fig2}.}

\label{fig4}
\end{figure}

In Fig. \ref{fig2}  we present the results of collapse dynamics with the first set of the rate parameters, e.g., $K_3^{(bb)} =  7.5\times 10^{-28}$ cm$^6$/s, $K_3^{(bf)} =  1.5\times 10^{-27}$ cm$^6$/s, for the binary mixture of 
20000 $^{164}$Dy atoms and 5000 $^{161}$Dy atoms. The collapse is initiated by jumping the interspecies scattering length $a_{bf}$ from 
$100a_0$ to $-100a_0$. In Fig. \ref{fig2} (a) we show the time evolution of the 
number of bosonic and fermionic atoms $N(t)/N(0)$ during collapse. 
Due to a rapid initial 
three-body loss during collapse, 
the number of both bosonic and fermionic atoms reduce with time and after the initial loss of a significant fraction of atoms the rate of loss of atoms is much reduced and a remnant condensate is formed which survives for a long period of time as in the collapse of a single-component nondipolar  BEC
\cite{donley}. To visualize the collapse dynamics closely we study the 3D isodensity plots of the condensates. In Figs. \ref{fig2} (b) and (c) we show the initial disk-shaped profiles of the bosonic and fermionic condensates. In Figs. \ref{fig2}
(d) $-$ (i), we illustrate the profiles of the bosonic and fermionic condensates at times $t\omega = 4$, 8 and 12. The evolutions of bosonic and fermionic profiles are distinct. In the radial plane the dipolar BEC develops maxima and minima due to dominant dipolar interaction. Similar maxima and minima were found to appear in binary dipolar BEC near the onset of instability due to increased dipolar interaction \cite{dipolarbin}. In Fig. \ref{fig2} we find that 
the bosons maintain smooth profiles during collapse and atom loss whereas   the fermions experience violent dynamics and pass through  unsmooth profiles.   In this case  a sequence of  collapse (with an increase in central density)
and explosion (with a decrease in central density) starts at about $t\omega = 4$.  Subsequently, after the repeated explosions small fermionic fragments 
scatter all around and the fermions occupy a larger region in space as can be seen in Fig. \ref{fig2} (i) at $t\omega=12$.

In Fig. \ref{fig3} we present the results of fermionic collapse in the 
binary mixture for the second set of rate parameters, e. g., $K_3^{(bb)} 
= 7.5\times 10^{-27}$ cm$^6$/s, $K_3^{(bf)} = 1.5\times 10^{-26}$ 
cm$^6$/s. The collapse is again initiated by jumping the interspecies 
scattering length $a_{bf}$ from $100a_0$ to $-100a_0.$ In Fig. 
\ref{fig3} (a) we show the time evolution of the number of atoms during 
collapse. The larger loss rates in this case compared to those used in 
Fig. \ref{fig2}, results in a more rapid loss of  atoms in 
the beginning and smaller remnant condensates in the end. In case of the 
collapse of a single-component nondipolar BEC,  it was also noted that a
rapid initial loss of atoms leads to a smaller remnant BEC \cite{donley,13611}. 
 The initial 
profiles of the condensates in this case are the same as in Figs. 
\ref{fig2} (b) and (c) for bosons and fermions, respectively. In this 
case, the dipolar BEC collapses 
towards the center and occupy a smaller central region after the onset 
of collapse.  After atom loss, the bosonic profile continues smooth. However, 
the mark of collapse and explosion is more prominent in the case of fermions.
The fermionic profile becomes unsmooth and small granules of matter is visible in this case.   

In Fig. \ref{fig4} we present the results of fermionic collapse in the 
binary mixture for the third set of rate parameters, e. g., $K_3^{(bb)} 
= 1.5\times 10^{-28}$ cm$^6$/s, $K_3^{(bf)} = 6\times 10^{-28}$ 
cm$^6$/s, for collapse  initiated by jumping the interspecies 
scattering length $a_{bf}$ from $100a_0$ to $-100a_0.$  This case corresponds to the smallest loss rates. Consequently, in this case the remnant states after long time are the largest in size as can be seen in Fig. \ref{fig4} (a).  The time evolution of the profiles are shown in Figs. \ref{fig4}
(b) $-$ (e) at times $t\omega =4$ and 8.  The profiles at $t\omega=4$ are particularly interesting: the BEC $^{164}$Dy has a Saturn ring type profile while the Fermi   $^{161}$Dy has a biconcave red-blood-cell type profile. 
Similar structures were observed in a disk-shaped   binary dipolar BEC on the verge of collapse and  are a clear manifestation of dipolar interactions 
\cite{dipolarbin}. The sign of collapse and explosion is visible in the Fermi 
profile at $t\omega =4$. At $t\omega=8$ the  explosion and fragmentation 
more clearly seen in both  profiles.

\begin{figure}
\begin{center}
\includegraphics[width=\linewidth]{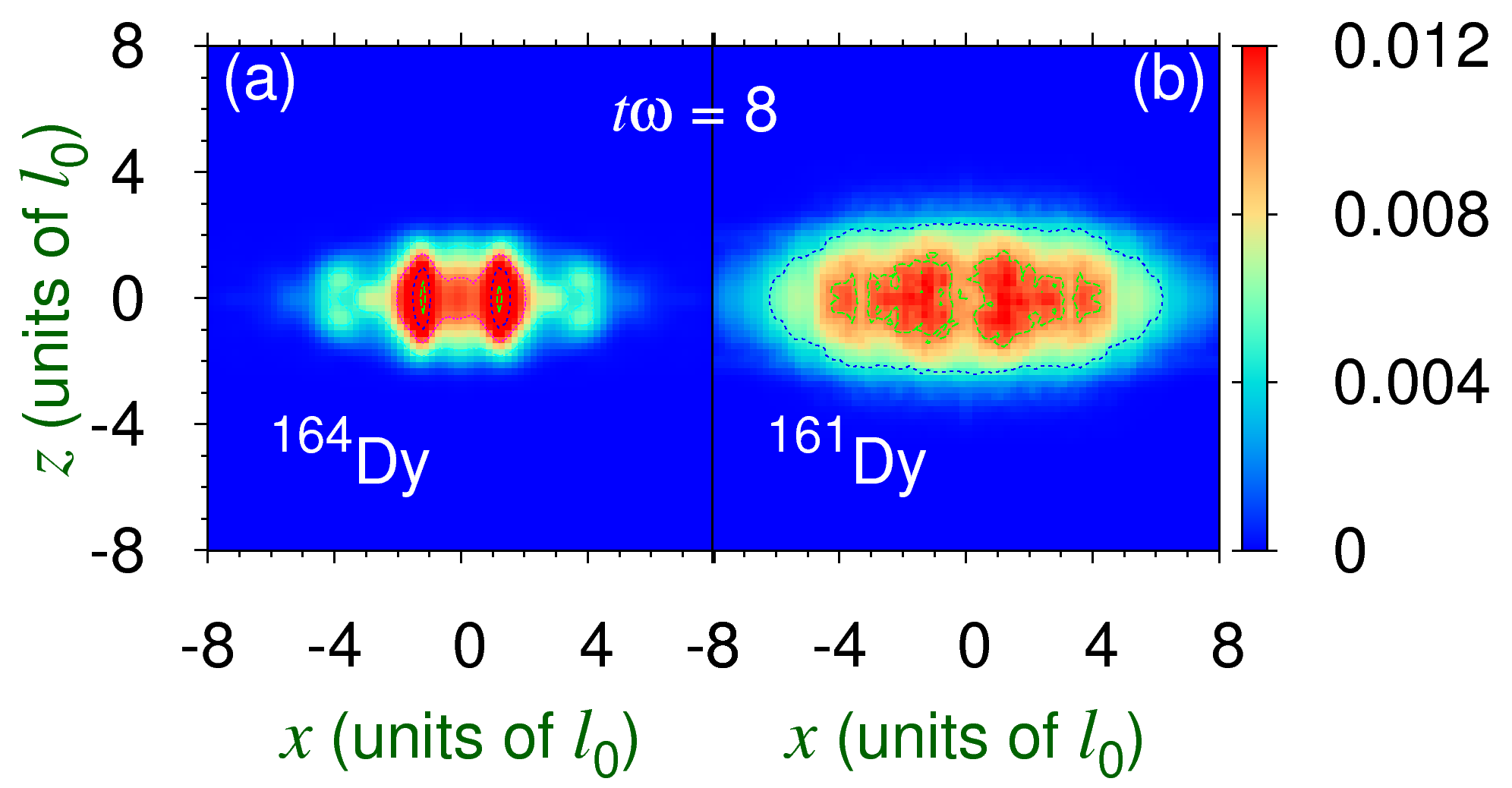} 
\includegraphics[width=\linewidth]{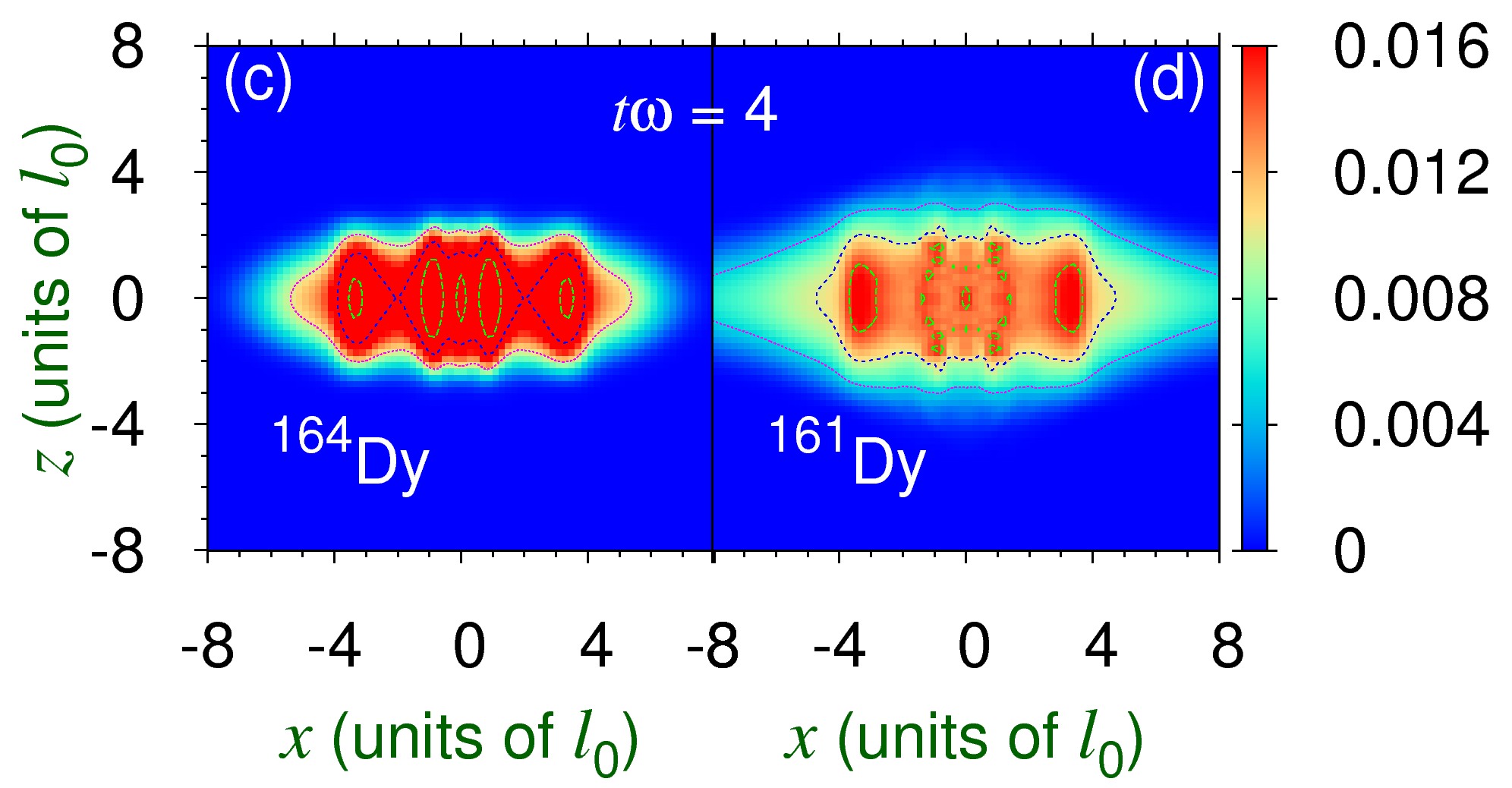} 
\end{center}

\caption{(Color online)   Effective 2D density of (a) $^{164}$Dy 
and (b) $^{161}$Dy atoms  
in the $x-z$ plane 
$|\phi(x,z,t)|^2\equiv \int dy |\phi({\bf r},t)| ^2$ in units of $l_0^{-2}$
at (dimensionless) time   
$t\omega =8$ for the dynamics shown in Fig. \ref{fig2} (f) and (g), respectively.
 The same 
of (c) $^{164}$Dy 
and (d) $^{161}$Dy atoms  
at time $t\omega =4$ for the dynamics shown in Fig. \ref{fig4} (b) and (c), respectively.}

\label{fig5}
\end{figure}

\begin{figure}
\begin{center}
\includegraphics[width=\linewidth]{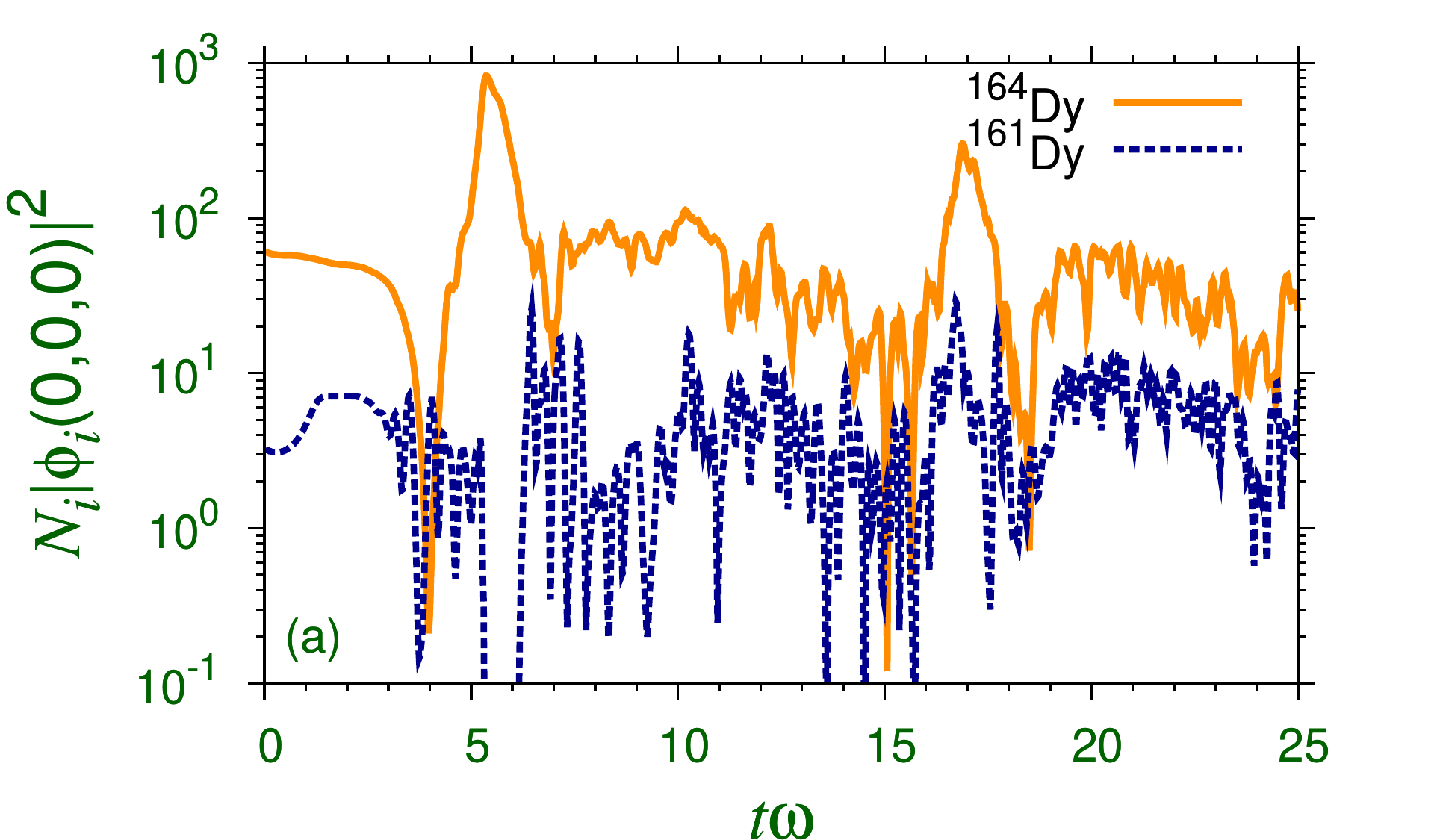} 
\includegraphics[width=\linewidth]{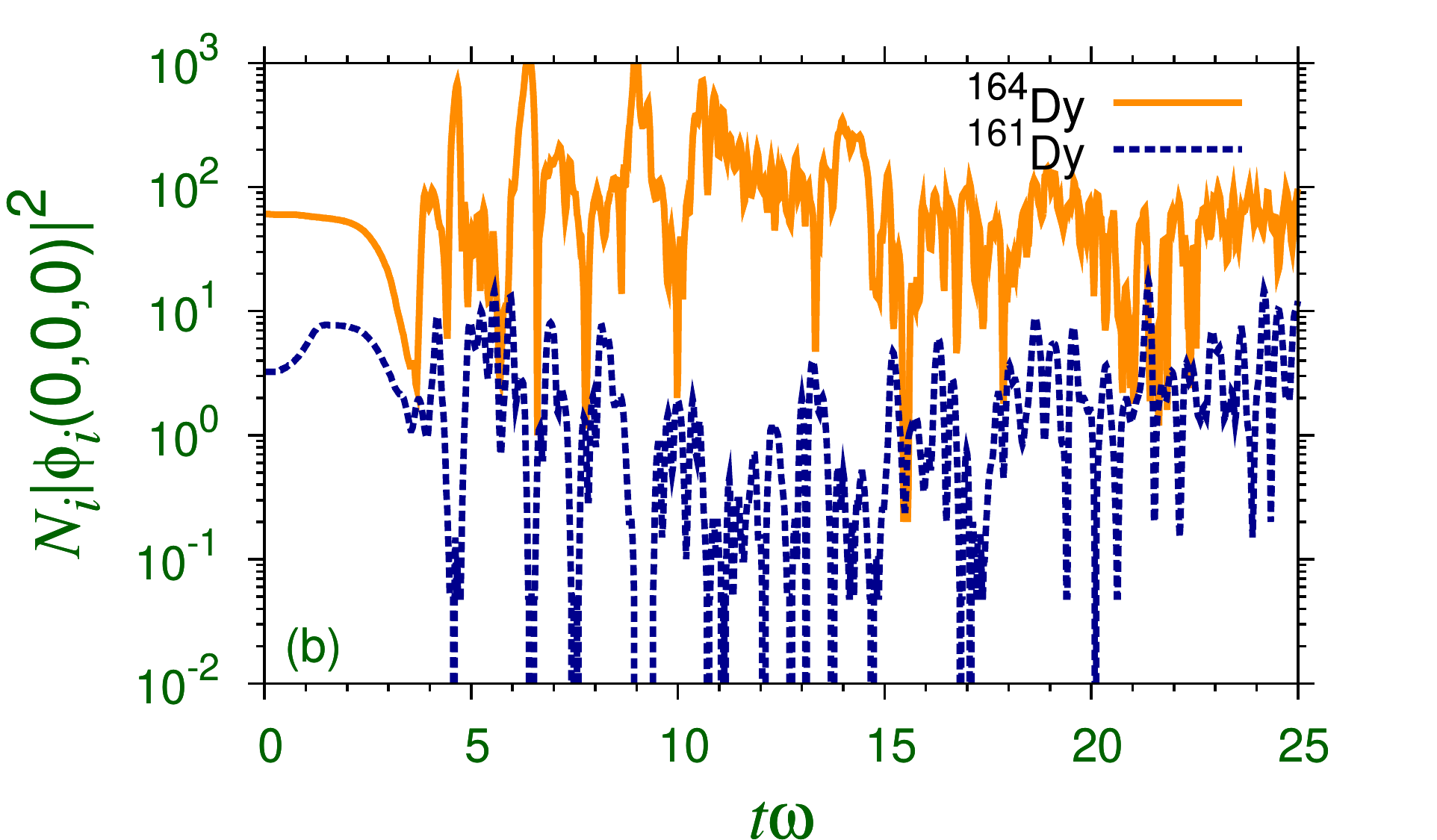} 
\end{center}

\caption{(Color online) Time evolution of the central density $N_i|\phi_i(0,0,0)|^2$ in units 
of $l_0^{-3}$
of the bosonic $^{164}$Dy and fermionic $^{161}$Dy atoms during 
collapse and explosion illustrated in (a) Fig. \ref{fig2} and (b) Fig. \ref{fig4}.}

\label{fig6}
\end{figure}

During collapse of the fermions the non-$S$ wave shape due to dipolar interaction appears in the fermionic profiles. To illustrate, 
we show in Figs. \ref{fig5} (a) and (b)
the contour plot of the effective 
2D density $|\phi(x,z,t)|^2 \equiv \int dy |\phi({\bf r},t)|^2$ in the $x-z$ plane for the bosonic $^{164}$Dy and fermionic $^{161}$Dy atoms, respectively,  
during the dynamics shown in Fig. \ref{fig2} (f) and (g)  at time $t \omega =8$. Both profiles for density are found to exhibit higher partial-wave shape due to dipolar interaction.
In Figs. \ref{fig5} (c) and (d) we show 
the contour plot of the effective 
2D density $|\phi(x,z,t)|^2$ 
 for the bosonic $^{164}$Dy and fermionic $^{161}$Dy atoms, respectively,  
during the dynamics shown in Fig. \ref{fig4}  (b) and (c)
at time $t \omega =4$.
The profile of the fermionic $^{161}$Dy atoms in Figs. 
\ref{fig5} (b) and (d)   is found to exhibit 
higher partial-wave shape due to dipolar interaction. The higher partial-wave shape in bosonic   $^{164}$Dy atoms also appears in Figs. \ref{fig5} (a) and 
(c). 
The saddle point \cite{saddle}
structures in density in the $x-z$ plane
in Figs. \ref{fig5} are manifestations of the dipolar
interaction. In these plots a minimum in density along the orthogonal 
$x$ direction coincides with a maximum in density in the polarization
$z$ direction thus creating a saddle point, which appears as a
clear manifestation of the saddle-shaped dipolar interaction \cite{saddle}. The  dipolar interaction leads to a effective trap with 
similar saddle shape \cite{saddle}. 
Such a density distribution is not possible in
the absence of dipolar interaction and also appears in a binary
dipolar BEC \cite{dipolarbin}. Similar non-$S$ wave density distribution 
was also noted in the study of collapse of a dipolar BEC of $^{52}$Cr atoms
\cite{12} due to a direct manifestation of dipolar interaction.

In Fig. \ref{fig6} (a) and (b) we show the evolution of central density 
$N_i|\phi_i(0,0,0)|^2$ of the bosonic $^{164}$Dy and fermionic $^{161}$Dy atoms during 
collapse and explosion illustrated in  Figs. \ref{fig2} and \ref{fig4}.
The repeated collapse to center and subsequent explosion appear as the 
rapid increase and decrease in the central density, respectively. 
The collapse and explosion in the fermions is found to be more vigorous
with rapid fluctuation in central density  
in the fermions than in the bosons in both cases.  The  reason 
is that the interspecies attractive contact interaction responsible for collapse acting  on 5000 fermions due to 20000 bosons is much stronger than the same acting on 20000 bosons due to 5000 fermions. Consequently, fluctuation in the central density in the dipolar BEC of $^{164}$Dy atoms 
is  smoother  than that in the fermionic 
 $^{161}$Dy atoms. 
 In the dynamics presented in Figs. \ref{fig6}, after a smooth evolution of the binary system till $t\omega\approx 3$, the cycle of collapse and explosion starts with a rapid fluctuation of  the central densities. Similar fluctuation of central densities was noted in the study of 
collapse of a nondipolar boson-fermion $^{87}$Rb-$^{40}$K mixture \cite{thcollbf}.

\section{summary and discussion}

Matter built of identical fermions is stable against collapse in the 
presence of isotropic short-range interaction due to a strong Pauli 
repulsion. On the other hand, bosons can exhibit collapse instability 
for attractive interactions \cite{donley}. The phenomenon of collapse 
and explosion in bosonic atoms has been observed \cite{donley} and 
studied \cite{13611,thcoll} in bosonic superfluids. However, it has been 
possible to observe collapse in fermions in the presence of bosons via 
an interspecies isotropic short-range interaction \cite{modugno}. It is 
also possible to have collapse instability in fermions in the presence 
of anisotropic long-range dipolar interaction \cite{dipolarb} operative 
in identical polarized fermions: the isotropic short-range interaction 
is forbidden in this case due to Pauli principle. Here we considered a 
more favorable situation of the collapse of fermions in a binary 
boson-fermion mixture including dipolar interaction as well as an 
attractive short-range boson-fermion interaction.

Because of recent experimental activities in a highly dipolar binary 
mixture of Dy isotopes \cite{boselev,122}, we studied fermionic collapse in the disk-shaped 
boson-fermion $^{164}$Dy-$^{161}$Dy mixture. The stability of the 
mixture was illustrated using phase diagrams of allowed critical number of 
bosonic $^{164}$Dy atoms in the binary mixture for different short-range 
interspecies and intraspecies interactions as well as for different trap 
aspect ratios. The collapse was started in a stable mixture by jumping 
the interspecies scattering length from a positive (repulsive) to 
negative (attractive) value using a Feshbach resonance \cite{fesh}.  
During collapse we include appropriate three-body loss rates via the 
formation of boson-boson and boson-fermion molecules. The evolution of 
the condensates during collapse was quantified by the evolution of 
respective atom numbers. The condensates first reduce in size rapidly by 
losing atoms via three-body loss and eventually become smaller remnant 
condensates which survive for a longer period of time. Such remnant 
condensates were observed in the collapse of nondipolar $^{85}$Rb BEC 
\cite{donley}.  We also identify anisotropic shapes of the condensates 
during collapse, which is a clear manifestation of dipolar interaction.  
The repeated collapse and explosion of the binary system were identified 
by
 rapidly oscillating central densities of the bosonic and fermionic 
atoms.  After repeated collapse and explosion, small fermionic fragments appear covering the whole region. All calculations in this 
paper are performed with realistic values of the parameters and could be 
verified in future experiments.


\acknowledgments
We thank FAPESP and CNPq (Brazil) for partial support.

\end{document}